\documentclass[pra,twocolumn,showpacs,eqsecnum,superscriptaddress]{revtex4}
\usepackage{graphicx}
\usepackage{verbatim}
\usepackage{color}
\usepackage{amsmath}
\usepackage{microtype}

\begin{document}

%Maxime's usual commands
\newcommand{\bra}[1]{\left\langle #1\right|}
\newcommand{\ket}[1]{\left|#1\right\rangle}
\newcommand{\braket}[2]{\left\langle #1|#2\right\rangle}
\newcommand{\com}[2]{\left[#1,#2\right]}
\newcommand{\braketop}[3]{\left\langle #1\left|#2\right|#3\right\rangle}
\newcommand{\mean}[1]{\left\langle #1 \right\rangle}
\newcommand{\trace}[2][]{{\rm Tr_{#1}}\left(#2\right)}
\newcommand{\ImaginaryPart}{{\rm Im}}
\newcommand{\RealPart}{{\rm Re}}
\newcommand{\elem}{\in}
\newcommand{\rp}{\right)}
\newcommand{\lp}{\left(}
\newcommand{\lcb}{\left\{}
\newcommand{\rcb}{\right\}}
\newcommand{\rsb}{\right]}
\newcommand{\lsb}{\left[}
\newcommand{\lbv}{\left|}
\newcommand{\rbv}{\right|}
\newcommand{\lvb}{\lbv}
\newcommand{\rvb}{\rbv}
\newcommand{\bs}{\boldsymbol}
\renewcommand{\inf}{\infty}
\newcommand{\order}[1]{{{\mathcal O}\lp#1\rp}}
\newcommand{\melem}[1]{_{#1}}
\newcommand{\iohbar}{\frac{-i}{\hbar}}
\newcommand{\ppm}{\stackrel{\raisebox{-1pt}[0pt][0pt]{$\scriptstyle(+)$}}{\raisebox{-4pt}[0pt][0pt]{$-$}}}
\newcommand{\pmp}{\stackrel{\raisebox{1pt}[0pt][0pt]{$\scriptstyle(-)$}}{\raisebox{-4pt}[0pt][0pt]{$+$}}}

%Constants
\newcommand{\wa}{{\omega_a}}
\renewcommand{\wr}{{\omega_r}}
\newcommand{\epm}{{\epsilon_m}}
\newcommand{\epc}{{\epsilon_c}}
\newcommand{\wm}{{\omega_m}}
\newcommand{\wc}{{\omega_c}}
\newcommand{\gd}{{\gamma_\downarrow}}
\newcommand{\gu}{{\gamma_\uparrow}}
\newcommand{\gphi}{{\gamma_\varphi}}
\newcommand{\gam}{{\gamma_1}}

\newcommand{\e}{e}
\newcommand{\g}{g}
\newcommand{\ag}{\alpha_\g}
\renewcommand{\ae}{\alpha_\e}
\newcommand{\aeg}{\alpha_{\e,\g}}
\newcommand{\am}{\beta}
\renewcommand{\ap}{\mu}
\renewcommand{\ne}{n_\e}
\renewcommand{\ng}{n_\g}
\renewcommand{\neg}{n_{\e,\g}}
\newcommand{\qrho}{{\rho}}
\newcommand{\crho}{{\varrho}}

%Operators
\newcommand{\ad}{{a^\dag}}
\renewcommand{\sp}{\sigma_+}
\newcommand{\sm}{\sigma_-}
\newcommand{\spm}{\sigma_\pm}
\newcommand{\sx}{\sigma_x}
\newcommand{\sy}{\sigma_y}
\newcommand{\sz}{\sigma_z}
\newcommand{\si}{\sigma_i}
\newcommand{\ip}{{I_+}}
\newcommand{\im}{{I_-}}
\newcommand{\ipm}{{I_\pm}}
\newcommand{\imp}{{I_\mp}}
\newcommand{\ada}{{\ad a}}
\newcommand{\Pe}{\Pi_\e}
\newcommand{\Pg}{\Pi_\g}
\newcommand{\Peg}{\Pi_{\e,\g}}
\newcommand{\Pa}{\Pi_\alpha}
\newcommand{\Dop}{D}

%Superoperators
\newcommand{\superop}[1]{{\mathcal #1}}
\newcommand{\sD}{{\superop{D}}}
\newcommand{\sL}{{\superop{L}}}
\newcommand{\sC}{{\superop{C}}}
\newcommand{\sT}{{\superop{T}}}
\newcommand{\sM}{{\superop{M}}}

%Unitary transforms
\newcommand{\tr}[1]{\mathbf{#1}}
\newcommand{\tU}{{\tr{U}}}
\newcommand{\tS}{{\tr{S}}}
\newcommand{\tD}{{\tr{D}}}
\newcommand{\tR}{{\tr{R}}}
\newcommand{\tP}{{\tr{P}}}
\newcommand{\tT}{{\tr{T}}}
\newcommand{\trans}[1]{^{#1}}

%Argument of Unitary transforms
\newcommand{\atr}[1]{{\mathbf #1}}
\newcommand{\atU}{{\atr{U}}}
\newcommand{\atS}{{\atr{S}}}
\newcommand{\atD}{{\atr{D}}}
\newcommand{\atR}{{\atr{R}}}
\newcommand{\atP}{{\atr{P}}}
\newcommand{\atT}{{\atr{T}}}

% Alexandre's commands
\newcommand{\aop}{a}
\newcommand{\eq}[1]{Eq.~(\ref{#1})}

%Jays commands
%standard commands
\newcommand{\nn}{\nonumber}
\newcommand{\nl}{\nn \\ &&}
\newcommand{\dg}{^\dagger}
\newcommand{\rt}[1]{\sqrt{#1}}
\renewcommand{\vec}[1]{\underset{\widetilde{}}{#1}}
\newcommand{\smallfrac}[2]{\mbox{$\frac{#1}{#2}$}}
\newcommand{\ito}{It\^o~}
\newcommand{\str}{Stratonovich~}
\newcommand{\sch}{Schr\"odinger~}
\newcommand{\schs}{Schr\"odinger's~}
\newcommand{\erf}[1]{Eq.~(\ref{#1})}
\newcommand{\erfs}[2]{Eqs.~(\ref{#1}) and (\ref{#2})}
\newcommand{\erft}[2]{Eqs.~(\ref{#1}) -- (\ref{#2})}
\newcommand{\szo}{\hat{\sigma}^z}
\newcommand{\sxo}{\hat{\sigma}^x}
\newcommand{\syo}{\hat{\sigma}^y}
\newcommand{\smo}{\hat{\sigma}^-}
\newcommand{\spo}{\hat{\sigma}^+}
\newcommand{\ano}{\hat{a}}
\newcommand{\cro}{\hat{a}\dg}
\newcommand{\Ho}{\hat{H}}
\newcommand{\Mso}[2]{{\cal M}_{#2}[{#1}]}

% Colours
\newcommand{\red}{\color[rgb]{0.8,0,0}}
\newcommand{\green}{\color[rgb]{0.0,0.6,0.0}}
\newcommand{\dkgrn}{\color[rgb]{0.0,0.4,0.0}} %0.6 0.1 0.3
\newcommand{\blu}{\color[rgb]{0,0,0.6}}
\newcommand{\blue}{\color[rgb]{0,0,0.6}}
\newcommand{\pur}{\color[rgb]{0.8,0,0.8}}
\newcommand{\blk}{\color{black}}

\title{Dispersive regime of circuit QED: photon-dependent qubit dephasing and relaxation rates.}
\date{\today}

\author{Maxime Boissonneault}
\affiliation{D\'epartement de Physique et Regroupement Qu\'eb\'ecois sur les Mat\'eriaux de Pointe, Universit\'e
de Sherbrooke, Sherbrooke, Qu\'ebec, Canada, J1K 2R1}
\author{J. M. Gambetta}
\affiliation{Institute for Quantum Computing and Department of Physics and Astronomy, University of Waterloo, Waterloo, Ontario, Canada, N2L 3G1}
\author{Alexandre Blais}
\affiliation{D\'epartement de Physique et Regroupement Qu\'eb\'ecois sur les Mat\'eriaux de Pointe, Universit\'e
de Sherbrooke, Sherbrooke, Qu\'ebec, Canada, J1K 2R1}

\begin{abstract}
Superconducting electrical circuits can be used to study the physics of cavity quantum electrodynamics (QED) in new regimes, therefore realizing {\em circuit} QED.  For quantum information processing and quantum optics, an interesting regime of circuit QED is the dispersive regime, where the detuning between the qubit transition frequency and the resonator frequency is much larger than the interaction strength.  In this paper, we investigate how non-linear corrections to the dispersive regime affect the measurement process.   We find that in the presence of pure qubit dephasing, photon population of the resonator used for the measurement of the qubit act as an effective heat bath, inducing incoherent relaxation and excitation of the qubit.  Measurement thus induces both dephasing and mixing of the qubit, something that can reduce the quantum non-demolition aspect of the readout. Using quantum trajectory theory, we show that this heat bath can induce quantum jumps in the qubit state and reduce the achievable signal-to-noise ratio of a homodyne measurement of the voltage.
\end{abstract}

\pacs{03.65.Yz, 42.50.Pq, 42.50.Lc, 74.50.+r, 03.65.Ta}

\maketitle

\section{Introduction} % (fold)
\label{sec:introduction}

Cavity quantum electrodynamics (QED) is a unique tool to study the interaction between light and matter at its most fundamental level~\cite{raimond:2001a,mabuchi:2002a}.  Most interesting is the regime of strong coupling where the frequency associated with light-matter interaction is greater than all relaxation rates~\cite{thompson:1992a,boca:2004a}.  For example, in this regime, experiments with very high-Q cavities have been able to resolve quantum jumps and time-resolved collapse of the cavity field~\cite{gleyzes:2007a,guerlin:2007a}.

With the strong coupling regime easily accessible, superconducting electrical circuits offer distinctive advantages to the study of light-matter interaction~\cite{buisson:2001a,marquardt:2001a,al-saidi:2001a,plastina:2003a,blais:2003d,you:2003a,yang:2003a,blais:2004a}, something which has been realized experimentally with charge~\cite{wallraff:2004a}, flux~\cite{chiorescu:2004a,johansson:2006}, and phase~\cite{sillanpaa:2007a} superconducting qubits.    Although the present work applies to all physical realizations of cavity or circuit QED, here we will focus on superconducting charge qubits coupled to a transmission line resonator~\cite{blais:2004a,wallraff:2004a}.  Because the qubit can be very strongly coupled to the transmission line in this system, it opens the possibility to study new regimes of cavity QED.  For example, the  strong dispersive regime was theoretically studied in Ref.~\cite{gambetta:2006a} and experimentally investigated in Ref.~\cite{schuster:2007a}.

In circuit QED, readout of the qubit is done by irradiating the resonator with photons at, or close to, the bare resonator frequency while the qubit is strongly detuned from the resonator.  Information about the state of the qubit is then encoded in the phase and amplitude of the field transmitted and reflected from the resonator.  In principle, increasing the amplitude of the measurement drive, hence the photon population of the resonator, should increase the rate at which information is gained about the qubit.  For example, in Ref.~\cite{blais:2004a,gambetta:2008a} it was estimated that by filling the resonator with the critical photon number $n_\mathrm{crit} = \Delta^2/4g^2$, where $\Delta$ is the frequency detuning between the qubit and the resonator, and $g$ is their interaction strength, one would reach signal-to-noise (SNR) ratios of $\sim 200$, even taking into account realistic amplifier noise.  Such SNR would easily lead to single-shot readout in this system~\cite{gambetta:2007a}.  However high SNR have not been experimentally observed. 

In previous these work, the conclusions for the SNR were obtained by analyzing the qubit-resonator Hamiltonian in the dispersive approximation, which is a perturbative expansion of the Jaynes-Cummings Hamiltonian to second order in $g/\Delta$.  However, this approximation fails as the number of photon in the resonator increases.  As $n_\mathrm{crit}$ is approached one should expect higher order terms in the perturbative expansion, with corresponding non-linearities, to be important.  While the dispersive approximation has been shown to be very accurate in understanding experimental results for circuit QED at moderate photon number population~\cite{wallraff:2004a,schuster:2005a,wallraff:2005a,gambetta:2006a,schuster:2007a,blais:2007a,houck:2007a,majer:2007a}, it should break down as the photon population is increased.  Understanding these corrections is important if we are to gain more insights in the measurement process and its effect on the qubit.

Here, this is done by pushing the dispersive approximation used in Ref.~\cite{blais:2004a,gambetta:2008a} to higher order.
These results not only apply to circuit QED but also to cavity QED and more generally to any physical situation where a two-level system is dressed by an oscillator.  Examples are quatronium~ \cite{siddiqi:2006a,boulant:2007a} or flux qubits~\cite{lupasu:2006a,lupascu:2007a,picot:2008a} coupled to bifurcating oscillators for readout purposes.    In particular, the authors of Ref.~\cite{picot:2008a} find that the qubit relaxation rate is strongly enhanced when the non-linear oscillator is in its high-amplitude state compared to its low-amplitude state, results which are at least qualitatively consistent with those presented here.

In Sec.~\ref{sec:dispersive_effects_on_the_hamiltonian} we find a unitary transformation that exactly diagonalizes the Jaynes-Cummings Hamiltonian. Expanding this to higher orders in $g/\Delta$ allows us to derive results which are valid for for higher photon numbers (but still less then $n_\mathrm{crit.}$). This transformation is then applied on the Hamiltonians describing coupling of the resonator and qubit to environmental degrees of freedom.  Taking advantage of the large separation in energy scale in the dispersive regime for the relevant qubit and resonator bath frequency, we obtain in Sec.~\ref{sec:dispersive_effect_on_the_master_equation} a Markovian Lindblad-type master equation for the system that takes into account higher order dispersive corrections and potential frequency-variations in the environmental spectral densities.

One of the most important results of this paper is obtained in Sec.~\ref{sec:effective_qubit_master_equation_eliminiation_of_the_cavity}.  There, building on Ref.~\cite{gambetta:2008a}, we eliminate the resonator degree of freedom from the resonator-qubit master equation found in Sec.~\ref{sec:dispersive_effect_on_the_master_equation} and derive an effective master equation for the qubit's reduced density operator. This is possible when $n<\{n_\kappa,n_\mathrm{crit.}\}$ where $n_\kappa$ is a new maximum photon number set by the cavity decay rate over the strength of the non-linearity. This effective master equation contains the measurement-induced dephasing found in Refs. \cite{schuster:2005a,gambetta:2006a,gambetta:2008a} and the novel effect of photon-number dependent qubit relaxation and dephasing rates. For example we show how finite photon population of the resonator acts as an effective heat bath on the qubit. This effective model is shown numerically to be very accurate in reproducing the dynamics of the full Jaynes-Cummings model. By comparing results obtained with the linear dispersive approximation, we find that non-linear effects become important even for photon occupation number significantly below $n_\mathrm{crit}$.  In Sec.~\ref{sec:second_order_effects}, the photon-dependent qubit mixing and dephasing rates are discussed in more details and the analytical results are compared to numerical calculations.

In Sec.~\ref{sec:dispersive_effects_on_the_stochastic_master_equation} a quantum trajectory equation describing the evolution of the qubit and resonator under homodyne measurement is obtained, as well as a reduced qubit quantum trajectory equation. This  is used to investigate the measurement, where we show that the achievable SNR is decreased substantially by the non-linear effects.
% (end)

\section{Circuit QED} % (fold)
\begin{figure}[tp]
\centering \includegraphics[width=0.99\linewidth]{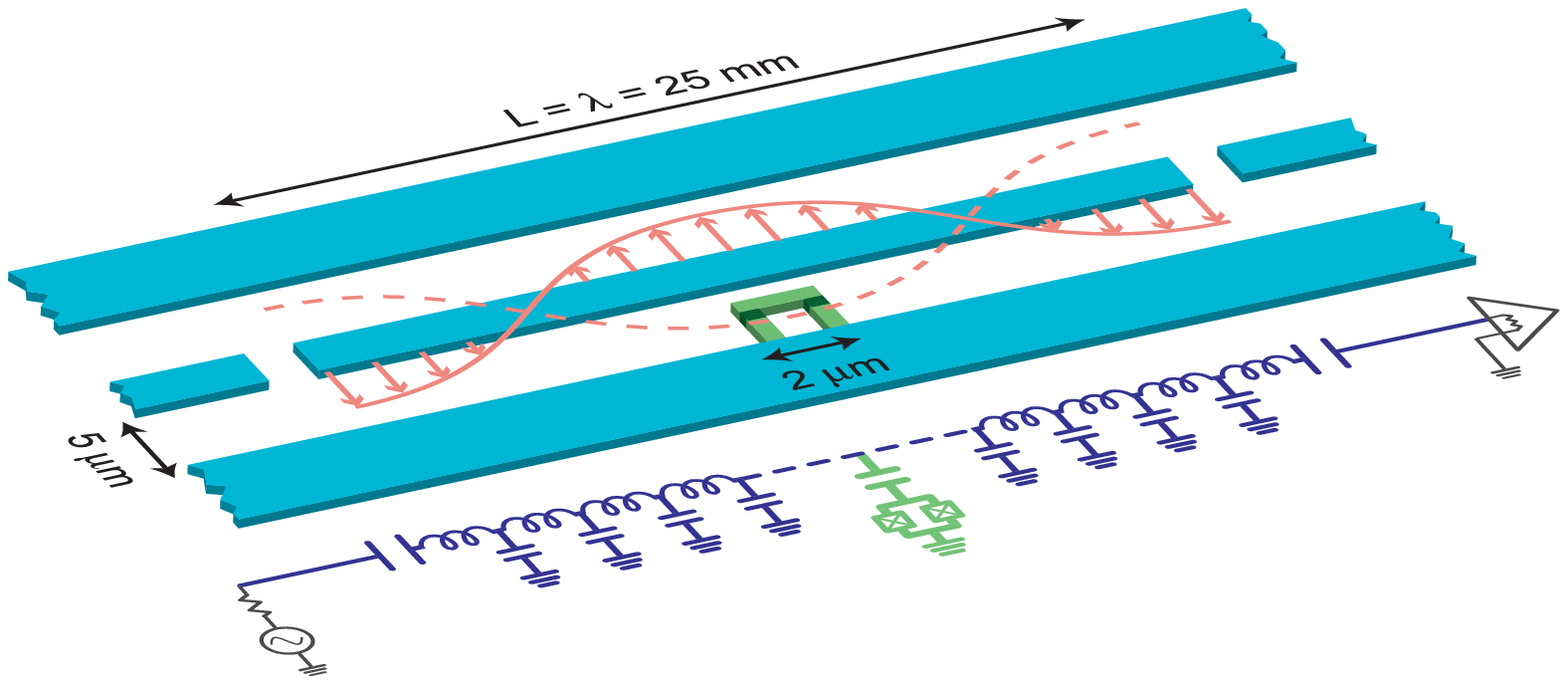}
\caption{(Color online) Schematic layout and lumped element version
of the  circuit QED implementation.  A superconducting charge qubit
(green) is fabricated inside a superconducting 1D transmission line
resonator (blue).} \label{fig_circuit}
\end{figure}

\label{sec:circuit_qed}
In circuit QED, a superconducting charge qubit is fabricated inside a transmission line resonator.  This system is illustrated in Fig.~\ref{fig_circuit}.  Focussing on a single mode of the resonator, the system Hamiltonian describing this circuit takes the Jaynes-Cummings form~\cite{blais:2004a}
\begin{equation}
	\label{eqn:HJC}
	H_s = H_0 + \hbar g\ip
\end{equation}
where we have defined
\begin{align}
	H_0 &= \hbar\wr \ad\aop + \hbar\wa \frac{\sz}{2} \\
	\ipm &= \ad \sm \pm a\sp.
\end{align}
In this expression, $\wr$ is the frequency of the mode of interest of the resonator, $\wa$ the qubit transition frequency and $g$ the qubit-resonator coupling.  The operators $a^{(\dag)}$ and $\spm$ are the creation and annihilation operators for the photon field and the qubit.

Logical operations and readout of the qubit can be achieved by applying a microwave signal on the input port of the resonator.  Choosing a frequency that is close to the resonator frequency $\wr$ corresponds to a readout of the qubit's state, while frequencies that are close to $\wa$ can be used to control the qubit~\cite{blais:2004a,gambetta:2006a,blais:2007a}.  This can be modeled by the Hamiltonian
\begin{equation}\label{eqn:H_drive}
	H_d = \sum_k \hbar\lp \epsilon_k(t) a^\dag e^{-i\omega_k t} + \epsilon^*_k(t) a e^{i\omega_kt}\rp.
\end{equation}
Taking into account that signals of different amplitude, frequency and phase can be sent simultaneously to the input port of the resonator,  $\epsilon_k(t)$ is the amplitude of the $k^\mathrm{th}$ drive and $\omega_k$ its frequency.  In this paper, we will be more particularly interested in looking at the effect of a measurement ($k=m$) drive and will have only this drive.

\subsection{Damping} % (fold)
\label{sub:damping}
The effect of coupling to environmental degrees of freedom can be described by the master equation
\cite{gardiner:2004b} 
\begin{align}\label{eqn:ExactMasterEquation}
	\dot\crho &= \iohbar[H,\crho] + \kappa \sD[a]\crho + \gamma_1 \sD[\sm]\crho + \frac{\gamma_\varphi}{2}\sD[\sz]\crho,
\end{align}
where $H$ is the total Hamiltonian of the system including drives
\begin{equation}\label{eqn:TotalHamiltonian}
	H = H_s + H_d
\end{equation}
and  $\sD[L]\crho = \lp 2 L\crho L^\dag - L^\dag L\crho - \crho L^\dag L\rp/2$.  In the above expression, $\kappa$ is the resonator rate of photon loss, $\gamma_1$ the qubit energy decay rate and $\gamma_\varphi$ is the qubit rate of pure dephasing.

This master equation is obtained in the Markov approximation which assumes that the spectral density of the environment is frequency-independent.  For high-quality factor systems, like high-Q transmission line resonators or (most) superconducting qubits, this approximation is accurate as the system is probing the environment in a very small frequency bandwidth.  As we will see in Sec.~\ref{sec:dispersive_effect_on_the_master_equation}, when going to the dispersive approximation, it can be important to take into account the frequency dependence of the environment. 

% subsection damping (end)
% section circuit_qed (end)

\section{Dispersive Effects on the Hamiltonian} % (fold)
\label{sec:dispersive_effects_on_the_hamiltonian}

In the limit that detunning between the cavity and the qubit is large, no energy is exchange.  In this situation, the interaction is said to be dispersive. In analyzing this interaction, it is convenient to diagonalize the Jaynes-Cummings Hamiltonian Eq.~\eqref{eqn:HJC} by using a unitary transformation.  In this transformed frame, the new effective qubit and photon operators are combinations of the bare qubit and photon operators.  In this sense, the qubit acquires ``a photon part" and vice versa.  This leads to, for example, the Purcell effect where a qubit can decay through the photon decay channel~\cite{houck:2008a,purcell:1946a}.  
% 
% In previous works~\cite{blais:2004a,gambetta:2006a,gambetta:2008a}, this diagonalization transformation was performed to first order in $g/(\wa - \wr)$.  These results are reviewed in the next section, before discussing exact diagonalization.

\subsection{Dispersive Jaynes-Cummings Hamiltonian: Linear regime} % (fold)
\label{sub:dispersive_jaynes_cummings_hamiltonian_linear_regime}

In the limit where $|\Delta|\equiv |\wa - \wr|\gg g$, the Jaynes-Cummings Hamiltonian~ \eqref{eqn:HJC} can be approximately diagonalized using the unitary transformation
\begin{equation}
	\tD_\mathrm{Linear} = e^{\lambda \im},
\end{equation}
with $\lambda=g/\Delta$ a small parameter.  Using the relation
\begin{equation}
e^{-\lambda X} H e^{\lambda X} = H + \lambda [H,X]  + \frac{\lambda^2}{2!}[[H,X],X]+\cdots 
\end{equation}
to second order in $\lambda$, it is simple to obtain the effective Hamiltonian describing the dispersive regime
\begin{equation}
	\begin{split}
		H_\mathrm{eff} &= \tD^\dag_\mathrm{Linear} H_s \tD_\mathrm{Linear} \\
		&= \hbar\wr \ad \aop + \hbar\lp \wa + 2g\lambda\lsb \ad \aop+\frac12\rsb \rp\frac{\sz}{2} + \mathcal{O}(\lambda^2).
	\end{split}	
	\label{eqn:ApproxHDisp}
\end{equation}

The qubit transition frequency is shifted by a quantity proportional to the photon population $2g\lambda\mean{\ada}$.  Alternatively, this shift can be seen as a qubit dependent pull of the resonator frequency $\omega_r \rightarrow \omega_r \pm g\lambda$.  
	
	As a result, shinning microwaves at the input port of the resonator at a frequency close to $\omega_r$ and measuring the transmitted signal using standard homodyne techniques serves as measurement of the qubit~\cite{blais:2004a,gambetta:2006a,schuster:2005a,wallraff:2005a,gambetta:2008a}.  In this approximation, this corresponds to a quantum non-demolition (QND) measurement of the qubit~\cite{gambetta:2006a,gambetta:2008a}. 
	
	For this measurement scheme, increasing the number of photons in the input beam increases the intensity of the output signal which would overcome the noise introduced by the amplifier. However the Hamiltonian~\eqref{eqn:ApproxHDisp} is only valid for a mean photon populations $n \ll n_{\rm crit.} = 1/4\lambda^2$ (as is shown by a Taylor expansion of the exact result Eq.~\eqref{eqn:DispersiveHamiltonianCompact}] which means we need to consider non-linear corrections to Eq.~\eqref{eqn:ApproxHDisp} to understand the dynamics as $n$ approaches $n_\mathrm{crit.}$.

\begin{figure}[pt]
	\centering
	\includegraphics[width=0.95\hsize]{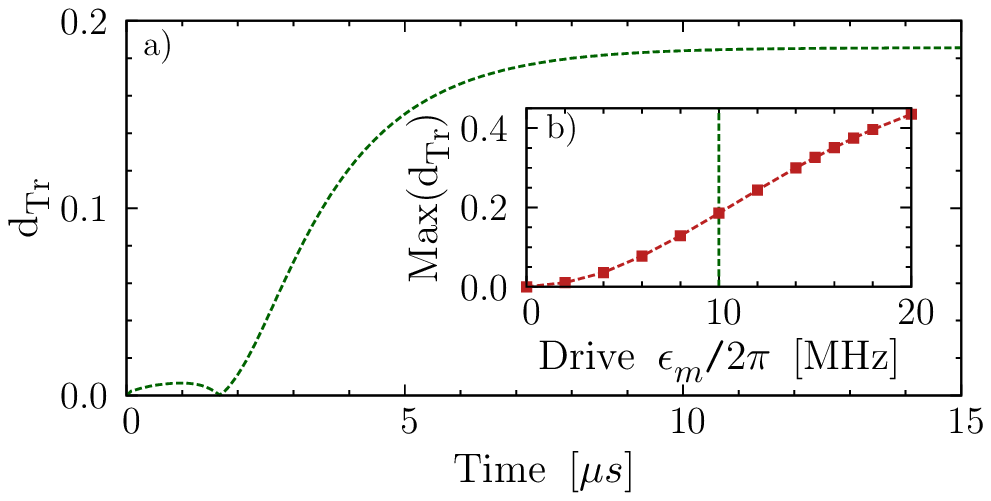}
	\caption{(Color online) Trace distance between the simulation of the full master equation and the dispersive linear model presented in Ref~\cite{gambetta:2008a} versus time. a) The initial state is $\ket{0}(\ket{\g}+\ket{\e})/\sqrt2$ and  $g/2\pi=50$MHz, $\Delta/2\pi=2$GHz, $n_{\rm crit.}=400$, $\kappa/2\pi=2.5$MHz, $\gamma_1/2\pi=0.1$MHz, $\gamma_\varphi/2\pi=0.3$MHz. A measurement drive of amplitude $\epm/2\pi=10$MHz of shape $0.5\epm (\tanh\lsb(t-\mu1)/\sigma\rsb+1)$ with $2\pi\mu_1 = 10.0\mu$s and $2\pi\sigma=10.0\mu$s is applied, corresponding to a mean of $\mean{\ada}\approx 34 \approx 0.08 n_{\rm crit.}$ photons. b) Maximum of the trace distance vs measurement amplitude. The vertical line indicates the measurement amplitude used in panel a).
	}
	\label{fig:dtr_justify}
\end{figure}

To see the breakdown of the linear approximation, we have numerically calculated the time-dependent evolution of the system master equation under measurement (see Fig.~\ref{fig:dtr_justify} for details) both in the linear dispersive approximation, using the approach described in Ref.~\cite{gambetta:2008a}, and with the full non-dispersive Jaynes-Cummings model.  To compare these results we plot the trace distance
\begin{equation}
	\begin{split}
		d_{Tr}(\rho_1,\crho_2) &= \frac12 \trace{|\rho_1 - \trace[r]{\crho_2}|} \\
		&= \sqrt{\sum_{i\in\{x,y,z\}}(\mean{\sigma_i}_1-\mean{\sigma_i}_2)^2},		
	\end{split}
\end{equation}
where $\rho_1$ is the reduced qubit density matrix found using the linear dispersive model presented in Ref~\cite{gambetta:2008a}, and $\trace[r]{\crho_2}$ is the trace over the resonator of the total density matrix of the system found by simulation of the complete master equation~\eqref{eqn:ExactMasterEquation}. This trace distance is the geometrical distance between two Bloch vectors, and ranges from  $0$ to $2$, with $0$ when the two qubit states are the same and $2$ when they are opposite on the Bloch sphere.

We plot, in figure~\ref{fig:dtr_justify}~a), the trace distance for a measurement amplitude which is slowly turned on to reach an amplitude corresponding to $\bar n \approx 0.08 n_{\rm crit.}$ photons in the resonator.  Even for this small number of measurement photons (compared to $n_{\rm crit.}$) the trace distance  is non-negligible which implies the breakdown of the dispersive approximation.  Moreover, we plot in figure~\ref{fig:dtr_justify}b) the maximum of the trace distance (over the simulation time), as a function of the maximum measurement amplitude $\epm$.  Clearly, the trace distance gets worse as the amplitude is increased. The importance of this effect  depends on the various parameters entering in the simulation but the results shown here are typical.  It is clear from these numerical results, it is important to take into account higher order terms in the dispersive approximation.
% subsection dispersive_jaynes_cummings_hamiltonian_linear_regime (end)

\subsection{Dispersive Jaynes-Cummings Hamiltonian: Exact transformation} % (fold)
\label{sub:dispersive_jaynes_cummings_hamiltonian_exact_transformation}

 Following the derivation presented in Appendix~\ref{sct:ExactDiag}, which is similar in spirit to the approach used in Refs~\cite{carbonaro:1979a,lo:1998a}, we find the unitary transformation that diagonalizes the Jaynes-Cummings Hamiltonian $H_s$
\begin{equation}
	\tD = e^{-\Lambda(N_q) \im},
	\label{eqn:D_transform}
\end{equation}
where 
\begin{align}
	\Lambda(N_q) &= -\frac{\arctan\lp2\lambda\sqrt{N_q}\rp}{2\sqrt{N_q}} \\
	\label{eqn:N_q}
	N_q &\equiv \ad\aop + \Pe.
\end{align}
$N_q$ is an operator representing the total number of excitations, and $\Pe$ is the projector on the qubit excited state. Applying this transformation to $H_s$ yields
\begin{equation}
	\label{eqn:DispersiveHamiltonianCompact}
	\begin{split}
		H_s\trans{\atD} &\equiv \tD^\dag H_s \tD\\ 
		&= \hbar\wr\ada + \hbar\wa\frac\sz2 - \frac{\hbar\Delta}{2} \lp 1 - \sqrt{1+4\lambda^2N_q} \rp \sz.
	\end{split}
\end{equation}
As it should, the eigenenergies of this Hamiltonian are the same as those presented in Ref.~\cite{blais:2004a} if $n$ is taken as the eigenvalues of $N_q$ and each eigenenergy is shifted by a constant $\hbar\wr$.

	In this basis, the qubit is dressed by the field.  As a result, qubit operators acquire photon part and similarly for field operators.  For example, under the transformation $\tD$, $\sz$ and $\ada$ become
\begin{align}\label{eqn:szS}
	\sz\trans{\atD} &= \sz\lp\frac{1}{\sqrt{1+4\lambda^2 N_q}}\rp - \frac{2\lambda}{\sqrt{1+4\lambda^2 N_q}}\ip \\
	\label{eqn:NS}
	(\ada)\trans{\atD} &= \ada + \frac{\sz}{2} + \frac{\lp\lambda\ip - \sz/2\rp}{\sqrt{1+4N_q\lambda^2}} .
\end{align} 
Both these operators now involve the off-diagonal operator $\ip$.

Expanding Eq.~\eqref{eqn:DispersiveHamiltonianCompact} to third order in $\lambda=g/\Delta$ [one order up from Eq.~\eqref{eqn:ApproxHDisp}] we find
\begin{equation}
	\label{eqn:H_s_D}
	\begin{split}
		H_s\trans{\atD} &\approx \hbar(\wr+\zeta)\ada + \hbar\lsb\wa+2\chi\lp \ada+\frac12\rp\rsb\frac\sz2 \\
		&\quad + \hbar \zeta (\ada)^2 \sz,
	\end{split}
\end{equation} 
where
\begin{equation}
\chi = g^2(1-\lambda^2)/\Delta
\end{equation} 
is the modified value of Lamb and Stark shift per photon.  In addition to a correction to these values, the third order expansion yields a squeezing term $(\ada)^2$ of amplitude $\zeta = -g^4/\Delta^3$. % This last term provides interesting physics and will be discussed in more detail in a later work.
% subsection dispersive_jaynes_cummings_hamiltonian_exact_transformation (end)

\subsection{The drive Hamiltonian under the exact transformation} % (fold)
\label{sub:the_drive_hamiltonian_under_the_exact_transformation}

It is important not only to transform $H_s$ but also the drive Hamiltonian $H_d$.  To do so, we first consider how  the qubit and field ladder operators are transformed under $\tD$. Contrary to $\sz$ and $N$, the transformation does not lead to a compact result.  To order $\mathcal{O}(\lambda^3)$ for $a$ and $\mathcal{O}(\lambda^2)$ for $\sm$, we find
\begin{align}
	\begin{split}\label{eq:aD}
	a\trans{\atD} &\approx a\lsb 1+\frac{\lambda^2\sz}{2}\rsb + \lambda\lsb 1-3\lambda^2\lp \ada+\frac12\rp\rsb\sm \\
	&\quad + \lambda^3 a^2\sp,
	\end{split}\\
	\label{eq:sigmaD}
	\sm\trans{\atD} &\approx \sm\lsb 1-\lambda^2\lp \ada+\frac12\rp\rsb + \lambda a\sz - \lambda^2 a^2\sp,
\end{align}
such that the drive Hamiltonian~Eq.~\eqref{eqn:H_drive} becomes 
\begin{equation}
	\label{eqn:H_d_D}
	\begin{split}
		H_d\trans{\atD} &= \sum_k \epsilon_k a^\dagger \lp 1+\frac{\lambda^2\sz}{2}\rp e^{-i\omega_kt} + \mathrm{h.c.} \\
		&\quad + \sum_k \epsilon_k \lambda\lsb 1-3\lambda^2\lp \ada+\frac12\rp\rsb\sp e^{-i\omega_kt} + \mathrm{h.c.}	
	\end{split}
\end{equation}
With $\omega_k\sim\omega_r$, the first line of the above equation is responsible for measurement of the qubit.   Due to the $\lambda^2$ term, the effective measurement drive strength is affected by the state of the qubit.  It will be slightly larger or smaller depending on the qubit being in its excited or ground state.  As will be shown later, this lead to small corrections to the ac-Stark shifted qubit transition frequency and measurement-induced dephasing rate.  Moreover, choosing $\omega_k\sim\omega_a$, one could take advantage of the second line of Eq.~\eqref{eqn:H_d_D} to coherently control the qubit.  Again due to a $\lambda^2$ term, the effective strength of this control drive will be modulated by the number of photons in the cavity.
% (end)

% section dispersive_effects_on_the_hamiltonian (end)

\section{Dispersive Effect on the Master Equation} % (fold)
\label{sec:dispersive_effect_on_the_master_equation}

To obtain a complete description of the system in the dispersive regime, we also need to apply the dispersive transformation to the bath-system coupling.   In principle, this can be done by transforming the operators entering the dissipative terms of the Lindblad master equation~\eqref{eqn:ExactMasterEquation}.  Once transformed, these terms will typically involve both qubit and field operators which correspond to probing the environment at different frequencies than the untransformed dissipative terms.  Since the master equation is obtained in the Markov approximation, this frequency information is lost.  

Here, we go beyond this approximation by rederiving the qubit-resonator master equation.  We first  apply the dispersive transformation on the system-bath Hamiltonian and then trace out the bath degrees of freedom to finally obtain a master equation in the dispersive frame.

\subsection{System-bath Hamiltonians}
%\subsection{Qubit relaxation and photon decay} % (fold)
%\label{sub:qubit_relaxation_and_photon_decay}

Energy damping of the resonator ($\kappa$) and of the qubit ($\gamma$) can be modelled  by coupling to baths of harmonic oscillators with free Hamiltonians~\cite{gardiner:2004b}
\begin{equation}
	\begin{split}
		H_{B\kappa} &= \hbar \int_0^\inf \omega b_\kappa^\dag(\omega) b_\kappa(\omega)d\omega \\
		H_{B\gamma} &= \hbar \int_0^\inf \omega b_\gamma^\dag(\omega) b_\gamma(\omega)d\omega,
	\end{split}
\end{equation}
where $b^\dag_{\kappa,\gamma}(\omega)$ and $b_{\kappa,\gamma}(\omega)$ respectively create and annihilate an excitation of frequency $\omega$ in the resonator or qubit bath. Coupling to these baths is described by~\cite{gardiner:2004b}
\begin{equation}
	\label{eq:bathcoupling}
	\begin{split}
		H_\kappa &= i\hbar\int_0^\inf \sqrt{d_\kappa(\omega)}\lp f^*_\kappa(\omega)b_\kappa^\dag(\omega) - \mathrm{h.c.}\rp (a+\ad)d\omega \\
		H_\gamma &= i\hbar\int_0^\inf \sqrt{d_\gamma(\omega)}\lp f^*_\gamma(\omega)b_\gamma^\dag(\omega) - \mathrm{h.c.}\rp \sx d\omega,
	\end{split}
\end{equation}
where $d_i(\omega)$ is the density of modes of bath $i$ and $f_i(\omega)$ represents the coupling strength of the mode of frequency $\omega$ to the resonator or the qubit.

% subsection qubit_relaxation_and_photon_decay (end)

%\subsection{Qubit dephasing} % (fold)
%\label{sub:qubit_dephasing}

Dephasing in the bare basis occurs due to slow fluctuations of the qubit transition frequency.  For example, in a superconducting charge qubit this is primarily caused by charge noise~\cite{astafiev:2004a,astafiev:2006a}.  Dephasing can be modeled by adding the Hamiltonian
\begin{equation}
	\label{eq:classical_dephasing}
	H_\varphi = \hbar\nu f_\varphi(t)\sz.
\end{equation}
In this expression, $f_\varphi(t)$ is a random function of time with zero mean and $\nu$ is characteristic of the magnitude of the coupling of the qubit to the fluctuations.  Defining $f_\varphi(t) = \int_{-\inf}^{\inf} f_\varphi(\omega) e^{i\omega t} d\omega$, $H_\varphi$ can be written in frequency space as
\begin{equation}
	\label{eq:classical_dephasing_omega}
	H_\varphi = \hbar\nu \sz \int_{-\inf}^{\inf} f_\varphi(\omega) e^{i\omega t} d\omega.
\end{equation}
% subsection qubit_dephasing (end)

\subsection{Dispersive master equation} % (fold)
\label{sub:the_dispersive_master_equation}

As shown in appendix~\ref{annsec:obtaining_the_dispersive_master_equation}, applying the dispersive transformation on the above system-bath Hamiltonians and integrating out the bath degrees of freedom leads to the master equation
\begin{equation}
	\label{eqn:ExactMasterEquation_D}
	\begin{split}
		\raisetag{16pt}
		\dot\crho\trans{\atD} &= -i[H_s\trans{\atD}+H_d\trans{\atD},\crho\trans{\atD}] \\
		&\quad + \kappa\sD[a(1+\lambda^2\sz/2)]\crho\trans{\atD} + \gamma_\kappa\sD[\sm]\crho\trans{\atD} \\
		&\quad + \gamma\sD\left[\sm\{1-\lambda^2(\ada + 1/2)\}\right]\crho\trans{\atD} + \kappa_\gamma \sD[a\sz]\crho\trans{\atD} \\
		&\quad + \gphi \sD[\sz\{1 - 2 \lambda^2 (\ada +1/2) \}]\crho\trans{\atD}/2 \\
		&\quad + \gamma_\Delta \sD[\ad\sm]\crho\trans{\atD} + \gamma_{-\Delta} \sD[a\sp]\crho\trans{\atD} \\
		&= \sL\trans{\atD}\crho\trans{\atD},
	\end{split}
\end{equation}
where we have defined the rates $\kappa = \kappa_r$, $\gamma_\kappa = \lambda^2 \kappa_a$, $\gamma = \gamma_a$, $\kappa_\gamma = \lambda^2 \gamma_r$ with
\begin{subequations}
	\label{eqn:rates}
	\begin{align}
		\kappa_p &= 2\pi d_\kappa(\omega_p) \lvb f_\kappa(\omega_p)\rvb^2 \\
		\gamma_p &= 2\pi d_\gamma(\omega_p) \lvb f_\gamma(\omega_p)\rvb^2 \\
		\gamma_\varphi &= 2\nu^2 S(\omega\rightarrow 0) \\
		\gamma_{\pm \Delta} &= 4\lambda^2 \nu^2 S(\pm\Delta).
	\end{align}
\end{subequations}
In obtaining these result, we have taken into account the fact that the spectral weight of the environment can be non-white.  As a result, although we obtain a Markovian master equation, the rates depend explicitly on the qubit and resonator environments at different frequencies.  As is explained in appendix~\ref{annsec:obtaining_the_dispersive_master_equation}, in obtaining these results it was assumed that noise at the various relevant frequencies are independent.  This assumption is valid if the noise is relatively weak and the various frequencies entering the expression for the rates  are well separated one from another.  For superconducting charge qubits which experimentally show long coherence times (up to $2~\mu$s~\cite{Schreier:2008a}) and in the dispersive regime (where $\Delta$, and thus the frequency separation, is ~$10$ GHz), the above model is accurate.

\section{Effective qubit master equation: Eliminiation of the Cavity} % (fold)
\label{sec:effective_qubit_master_equation_eliminiation_of_the_cavity}

In this section, we eliminate the resonator degree of freedom from \eq{eqn:ExactMasterEquation_D} to obtain a master equation for the reduced qubit density matrix in the dispersive frame. Building on Ref.~\cite{gambetta:2008a}, this is done by first moving to a rotating frame for the cavity, and then using a polaron-type transformation to displace the cavity field back to the vacuum. From this frame it is possible to consider only the two classical fields $\ae$ and $\ag$. These fields correspond to the average value $\mean{a}$ of the cavity field if the qubit is in the excited or ground state.

The resulting master equation is valid as long as the resonator state does not deviate too much from a superposition of coherent state. This can be formalized by two requirements. The first is
\begin{equation}
	n \ll n_\kappa=\frac{\kappa}{|\zeta|}	
\end{equation}
where $n_\kappa$ is the ratio of the rate $\zeta$ at which the non-linearity is squeezing the resonator state and the rate $\kappa$ at which these deviations are taken back to coherents state by damping. The second requirement is
\begin{equation}
	\gd,\gu \ll \kappa,
\end{equation}
where $\gd$ and $\gu$ are given in Eqs~\eqref{eqn:Gamma_Down} and \eqref{eqn:Gamma_Up} and are the rates at which the superposition of the coherent states $\ae$ and $\ag$ are getting mixed.  This condition implies that the rate at which the superposition of the coherent states gets mixed is much slower than the rate of photon loss.

As derived in appendix~\ref{sec:the_polaron_transformation}, in a frame rotating at $\omega_m$ for the resonator, the effective qubit master equation is
{
\begin{equation}
	\label{eqn:ReducedMasterEquation}
	\begin{split}
		\dot\qrho\trans{\atD} &= -i\frac{\wa\trans{\atD}}{2}\com{\sz}{\qrho\trans{\atD}} + \frac{\gphi_{\rm eff}}{2}\sD[\sz]\qrho\trans{\atD} \\
		& \quad  + \gd\sD[\sm]\qrho\trans{\atD} + \gu\sD[\sp]\qrho\trans{\atD}, 	
	\end{split}
\end{equation}
where $\qrho = \trace[r]{\crho}$ is the reduced density matrix of the qubit.   The parameters introduced in this master equation are 
\begin{align}
	\begin{split}
		\raisetag{16pt}
		\label{eqn:Delta_acDR}
		\wa\trans{\atD} = & \wa' + 2\lsb\chi+\zeta(1+\ne+\ng)\rsb \RealPart[\ag\ae^*] - \zeta(\ng^2+\ne^2) \\
		&+ \lambda^2\RealPart[\epm\ap^*] - \frac{\gamma_{-\Delta} + \gamma_\Delta - \lambda^2\gamma}{2} \ImaginaryPart[\ag\ae^*]
	\end{split}\\
	\begin{split}
		\raisetag{16pt}
		\label{eqn:Gamma_Phieff}
		\gphi_{\rm eff} 
		=& \gphi\lsb1-\frac{\lp\ne+\ng+1\rp}{2n_\mathrm{crit.}}\rsb + \Gamma_\mathrm{d}
	\end{split}\\
	\begin{split}
		\raisetag{16pt}
		\label{eqn:Gamma_d}
		\Gamma_\mathrm{d}
		=& 2[\chi+\zeta(1+\ne+\ng)]\ImaginaryPart[\ag\ae^*]+ \lambda^2\ImaginaryPart[\epm\beta^*] \\
		& +\frac{\RealPart[(\gamma_{-\Delta}\ag - \gamma_\Delta\ae+\gamma\lambda^2\ae)\beta^*]}{2}
	\end{split}\\
	\begin{split}
		\label{eqn:Gamma_Down}
		\gd =& \gam\lsb1-\frac{\lp\ne+\frac12\rp}{2n_\mathrm{crit.}}\rsb + \gamma_\kappa + \gamma_{\Delta}(\ne+1)
	\end{split}\\
	\begin{split}
		\label{eqn:Gamma_Up}
		\gu =& \gamma_{-\Delta} \ng
	\end{split}
\end{align}
where the classical parts of the field $\ag$ and $\ae$ (pointer states) satisfy $(i\in\{e,g\})$
\begin{equation}
	\begin{split}
		\label{eqn:Condition_alphas}
		\dot\alpha_j 
			&= -i\epm\lp1 \pm \frac{\lambda^2}{2}\rp \\
			&\quad - i\lcb\Delta_{rm}' \pm \lsb\chi+2\zeta\lp n_j+\frac12\rp\rsb\rcb\alpha_j \\
			&\quad - \frac{\kappa(1\pm\lambda^2)+\kappa_\gamma+\gamma_{\pm\Delta}-\delta_{j,e}\gamma\lambda^2}{2}\alpha_j,
	\end{split}
\end{equation}
with the top sign for $j=e$ and the bottom sign for $j=g$, and with $\Delta_{rm}' = \Delta_{rm} + \zeta$, where $\Delta_{rm}=\omega_r-\omega_m$. In this expression, $\delta_{j,e}$ is the Kronecker delta. Following the notation of Ref.~\cite{gambetta:2008a}, we have used
\begin{equation} \label{eqn:beta_mu}
\beta = \ae-\ag, \qquad \mu = \ae+\ag.
\end{equation}
Moreover,  $\neg = \lvb\aeg\rvb^2$ is the number of photons in the cavity when the qubit is in the ground or excited state. With $\lambda^2 = 0$, the results of Ref~\cite{gambetta:2008a} are correctly recovered.

We now turn to a physically motivated description of these results.  First, $\wa\trans{\atD}$ is the qubit transition frequency with  $\wa' = \omega_a + \chi$ being the and Lamb shifted qubit frequency. The remaning terms are the ac-Stark shift. Then, Eq.~\eqref{eqn:Gamma_Phieff} is the qubit's pure dephasing rate.  The first term is the bare pure dephasing rate, which now depends on the photon number $n_e+n_g$ because of the dressing of the qubit by the field.  Interestingly, this rate {\em decreases} with photon population as dressing increases with photon number and the photons are unaffected by qubit dephasing. This rate always remains positive as $n_e$ and $n_g$ must always be smaller than $n_\mathrm{crit}$.

The term $\Gamma_\mathrm{d}$, which is defined in Eq.~\eqref{eqn:Gamma_d} is measurement-induced dephasing. The first term comes from information about the qubit state contained in the frequency dependence of the pointer states [see Eq. \eqref{eqn:Condition_alphas}], the second term is the qubit information encoded into the driving part of the pointer states [see Eq. \eqref{eqn:H_d_D}], and the last  one is qubit information encoded into the decay rates of the pointer states. As in previous work \cite{gambetta:2006a,gambetta:2008a}, this decay rate can be negative.  This is due to the recurrence of the qubit coherence which physically comes from the information about the qubit state which was lost into the cavity being transfered back into the qubit.  Positive constraints on the master equation bound how negative this rate can be but since our model is based on a physical model which is positive in the enlarged cavity-qubit space these negative rates will never lead to an unphysical state.

Eq.~\eqref{eqn:Gamma_Down} represents the effective qubit decay rate.  It's main contribution is proportional to $\gamma_1$ and, again, is reduced by dressing (but can never be negative).  The second term $\gamma_\kappa \sim \lambda^2\kappa$ is the Purcell effect which corresponds to qubit decay through the photon loss channel \cite{houck:2008a}. The last term of $\gd$ and $\gu$ are particularly interesting.  They describe, respectively, additional relaxation and excitation of the qubit due to the photons populating the resonator.  

{For the remaining of this section, we consider a purely white noise approximation, and a measurement drive at the resonator frequency, which ensures $\bar n = (\ne+\ng)/2 \approx \ne \approx \ng$. } As a result of these contributions, photons injected in the resonator for the measurement appear to the qubit as a heat bath of temperature $T=(\hbar\omega_r/k_\mathrm{B})/\log(1+1/\bar n)$.  This effective temperature depends on the measurement drive amplitude and frequency.  Under measurement, the qubit therefore suffers from additional mixing, something which can  reduce the quantum non-demolition aspect of the readout.  From Eq.~\eqref{eqn:rates}, we can write $\gamma_{\uparrow/\downarrow} = \gamma_{\pm\Delta} \bar n \sim 2\lambda^2\gphi  \bar n$ and these mixing rates are therefore due to both photon population and pure qubit dephasing.   This can be understood in the following way.  Let's assume the photon to initially be in the uniform superposition $(\ket{g}+\ket{e})/\sqrt{2}$ and the resonator in the vacuum state $\ket{0}$ (measurement drive is initially off).  This initial state is schematically illustrated by the light gray dots in Fig.~\ref{fig:mixing_ladder}.  When the measurement drive is turned on, the photon population increases to reach a poisson distribution centered about an average value $n$.  This is schematically illustrated by the black dots in Fig.~\ref{fig:mixing_ladder}.  Because of the qubit-resonator coupling, the state $\ket{e,n}$ acquires a component $\ket{g,n+1}$ and, likewise, $\ket{g,n}$ acquires a $\ket{e,n-1}$ component (illustrated by the wiggly arrows).  The amplitude of this qubit-resonator coherent `mixing' increases with photon population as $\sim\lambda\sqrt{n}$.  As illustrated in Fig.~7 of Ref.~\cite{blais:2004a}, in the absence of dephasing this mixing is completely coherently undone once the measurement drive is turned off leading to a QND measurement.  However, in the presence of pure qubit dephasing, the phase coherence in the qubit-resonator dressed states can be lost, leading to effective downward and upward incoherent transitions between the qubit states.  Given this, one should expect this rate to be proportional to the square of the qubit-resonator mixing amplitude $\lambda\sqrt{n}$ and to the dephasing rate, the result obtained in Eqs.~\eqref{eqn:Gamma_Down} and \eqref{eqn:Gamma_Up}.  

\begin{figure}[tp]
	\includegraphics[width=0.6\linewidth]{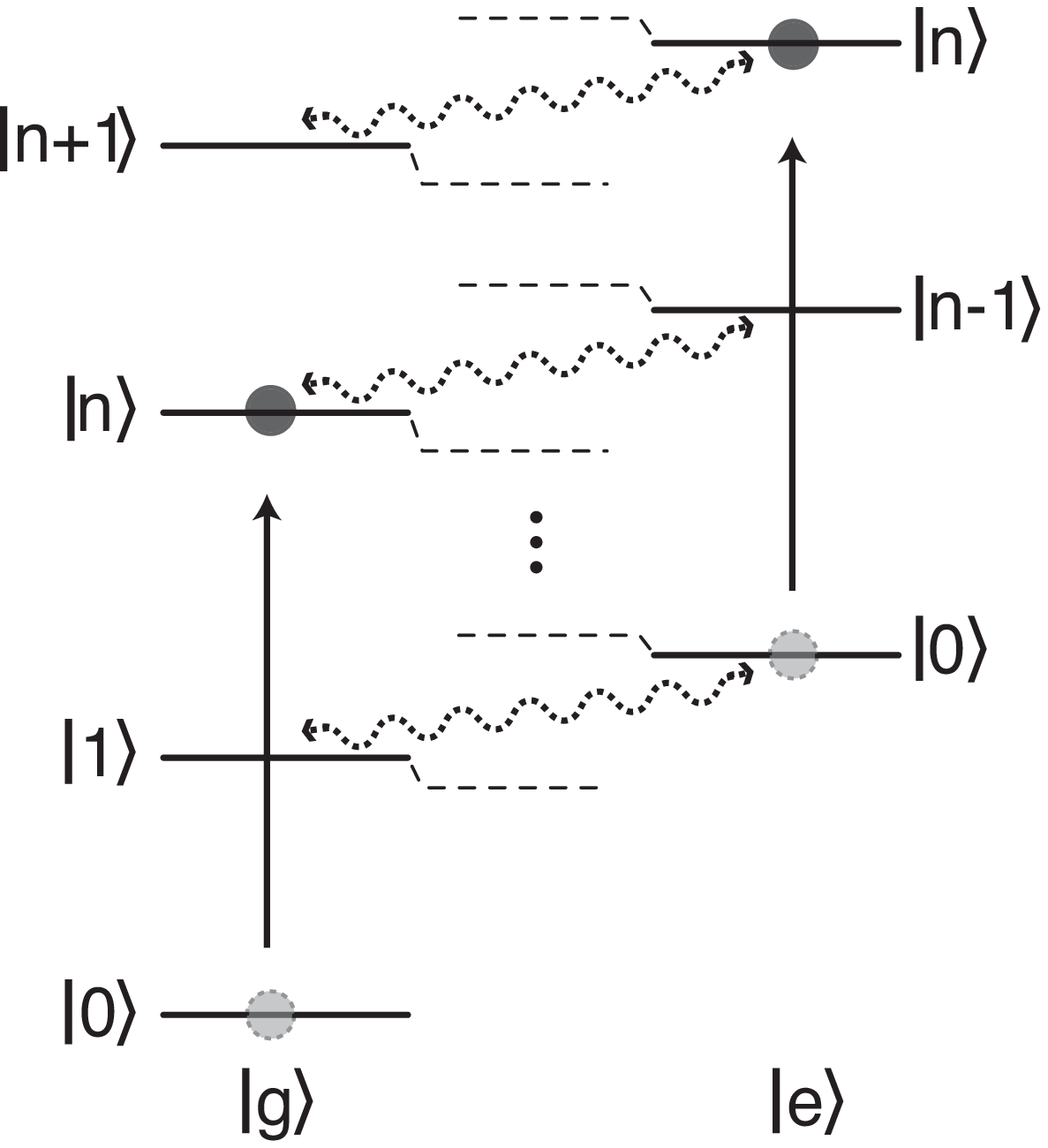}
	\caption{Dispersive energy diagram.  The full line represent the bare states while the straight dashed lines represent the qubit-resonator dressed states.}
	\label{fig:mixing_ladder}
\end{figure}

In summary, the rates $\gu$  and $\gd$ are due to dressing, by the resonator field, of the qubit operator $\sz{}$ causing dephasing in the bare basis.  We will therefore refer to this as dressed dephasing.

\subsection{Numerical comparision with the full master equation} % (fold)
\label{sub:comparision_with_the_full_master_equation}

To verify the validity of the previous results, we have done extensive numerical calculations in the limit $n\ll n_\kappa$ to compare results obtained from the reduced master equation~\eqref{eqn:ReducedMasterEquation} to those obtained from the qubit-resonator master equation~\eqref{eqn:ExactMasterEquation}.  The results obtained from \eq{eqn:ReducedMasterEquation} are also compared to those obtained from the linear approximation of Ref.~\cite{gambetta:2008a}.  From this latter comparison, it will be apparent that the non-linear model obtained here is much more accurate, while adding essentially no additional complexity in numerical simulation. 

\begin{figure}
	\centering
	\includegraphics[width=0.95\hsize]{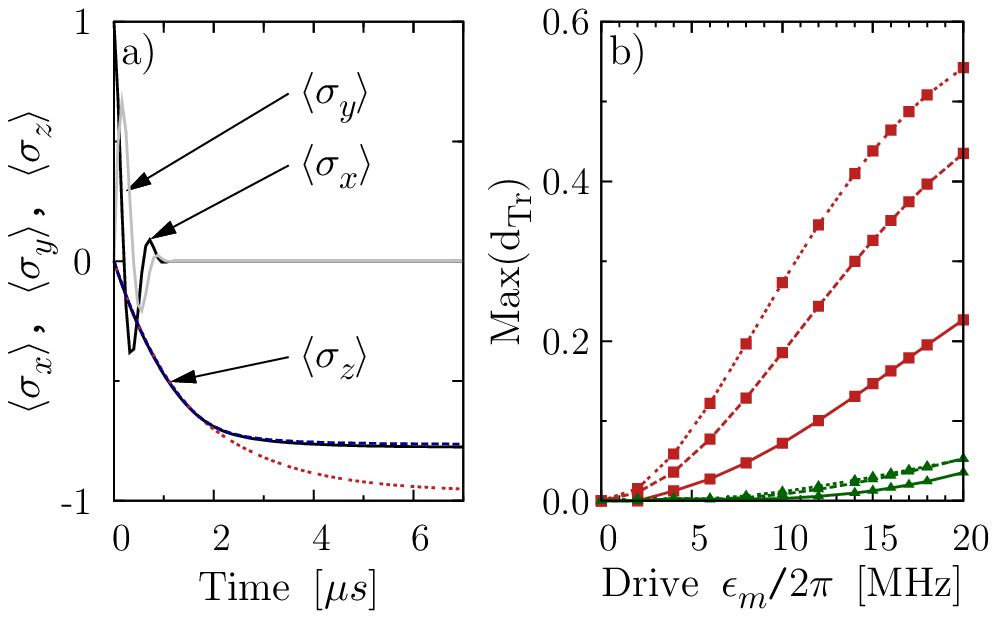}
	\caption{(Color online) Comparison between the exact master equation~\eqref{eqn:ExactMasterEquation} and the model~\eqref{eqn:ReducedMasterEquation}. a) A typical time evolution of $\sx$ (full black), $\sy$ (full grey) and $\sz$ for the exact result (full black), the non-linear (dashed blue) and the linear (dotted red) models. The parameters and the initial state are the same as Fig.~\ref{fig:dtr_justify}. 	b) Maximum of the trace distance for the linear (red squares) and the non-linear (green triangles) models, for $\gamma_\varphi/2\pi=0.5,0.3,0.1$MHz (dotted, dashed, full lines). }
	\label{fig:Full_DL_DC}
\end{figure}

Figure~\ref{fig:Full_DL_DC}a) presents a typical time evolution of the qubit as obtained by the numerical integration of the full master equation~\eqref{eqn:ExactMasterEquation} [full black line], the reduced model~\eq{eqn:ReducedMasterEquation} [dashed blue line] and the linear model of  Ref.~\cite{gambetta:2008a} [dotted red line]. The time evolution of $\mean{\sx}$ and $\mean{\sy}$ obtained from these three models are indistinguishable.  However, because the linear model does not capture dressed dephasing, only the non-linear model reproduces the correct equilibrium value of $\mean{\sz}$  

We note that the numerical results obtained using the full master equation have been time-averaged to get rid of small amplitude fast oscillations.  These oscillations are not contained in the effective models because of the various rotating-wave approximations that have been performed analytically.  Experimentally, this averaging is effectively performed due to the finite bandwidth of measurement apparatus.   Moreover, for simplicity, for the numerical results, a white noise spectrum was assumed.  We have therefore taken $\gamma_\kappa = \lambda^2\kappa$, $\kappa_\gamma = \lambda^2\gam$ and $\gamma_{\pm\Delta} = 2\lambda^2\gphi$ throughout this section.

Figure~\ref{fig:Full_DL_DC}b) shows the maximum of the trace distance (over time) as a function of measurement power for three values of the pure dephasing rate $\gamma_\varphi$. The red curves with square dots are the trace distances between the full and the linear reduced model, while the green curves with triangle dots are the trace distances between the full and the non-linear reduced model. Unsurprisingly, as the measurement power is increased, the trace distance between the reduced models and the exact solution increases.  As shown the three different curves for both models, the distance also increases as the dephasing rate is increased.    However, the non-linear model obtained here is clearly much more accurate than the linear one, it captures the physics of dressed-dephasing. The non-linear model also shows  much less variation in the trace distance with dephasing rate.  It is worth pointing out that the maximum measurement power used in Figure~\ref{fig:Full_DL_DC}b) corresponds to a very conservative photon population of the resonator $n\approx 0.4 n_{\rm crit}$, much lower than the critical photon number where non-linear effects were thought to become important~\cite{blais:2004a,gambetta:2008a}.

The effective model developed in this section is both accurate and much less demanding numerically than the full numerical integration of the qubit-resonator Hamiltonian. It should therefore be a useful tool to study the dispersive regime of circuit and cavity QED.
% subsection comparision_with_the_full_master_equation (end)

% section effective_qubit_master_equation_eliminiation_of_the_cavity (end)

\section{Qubit population and effective damping rate} % (fold)
\label{sec:second_order_effects}
In this section, we focus on the dependance of the qubit mixing rate on photon population and dephasing rate, and on the steady-state value of $\mean{\sz}$.  These two quantities could be measured experimentally as a test of the present model.

\subsection{Photon number dependant qubit decay rate} % (fold)
\label{sub:cavity_dependant_qubit_decay_rate}
A remarkable feature of the non-linear model is that the qubit up, down and dephasing rates depend on the photon population.  In particular, the effective qubit mixing rate is given by {
\begin{equation}
	\label{eqn:Gamma1Eff}
	\begin{split}
		\gamma_{\mathrm{eff}}(n_{\e s},n_{\g s}) &= \gamma_{\downarrow}(n_{\e s}) + \gamma_{\uparrow}(n_{\g s}) \\
		&= \gam\lsb 1-2\lambda^2\lp n_{\e s}+\frac12\rp\rsb + \gamma_\kappa \\
		&\quad + \gamma_\Delta(n_{\e s}+1) + \gamma_{-\Delta} n_{\g s},
	\end{split}
\end{equation}
where $n_{i s}=\lvb\alpha_{i s}\rvb^2$ are understood as the steady-state solutions of~\eq{eqn:Condition_alphas}. Interestingly, in the situation where $n_{\e s} \approx n_{\g s}$ and for white noise, } such that $\gamma_\Delta = \gamma_{-\Delta} = 2\lambda^2\gphi$, if $\gamma_\varphi < \gamma_1/2$, then increasing photon population leads to a {\em decrease} of the effective mixing rate.  On the other hand, if  $\gamma_\varphi > \gamma_1/2$, increasing photon population leads to an increase of the mixing rate.  This is again a consequence of dressing of the qubit by the photon field.

% subsection cavity_dependant_qubit_decay_rate (end)

\subsection{Measurement-induced heat bath} % (fold)
\label{sub:measurement_induced_heat_bath}
From the reduced qubit master equation~\eqref{eqn:ReducedMasterEquation}, the steady-state value of $\mean{\sz}$ can be expressed as
\begin{equation}
	\label{eqn:mean_sz_s}
	\mean{\sz}_s = -\frac{\gd-\gu}{\gd+\gu} = -1 + \frac{2 \gu(n_{\g s})}{ \gamma_{\rm eff}(n_{\e s},n_{\g s})}.
\end{equation}
While the linear model would predict $\mean{\sz}_s = -1$, the second order term $\gamma_{-\Delta}$ causes a deviation of $\mean{\sz}_s$ from this value which increases with $n_{\g s}$. This deviation is indicative of the breakdown of the QND aspect of the qubit measurement.

When comparing expectation values, it is of course important to compare expressions computed in the same basis.  As a result, it is useful to transform~\eqref{eqn:mean_sz_s} to the bare basis~\footnote{When computing the expectation value of an operator in a transformed basis, the transformation that was applied to the state vector must also be applied to the operator in order to get the expectation value in the non-transformed basis: $\langle A\rangle= \trace{A\rho} = \trace{\tD A \tD^\dag \tD \rho \tD^\dag}$.}.  This is done by applying  the dispersive transformation to $\sz$, see Eq.~\eqref{eqn:szS}, from which we obtain
\begin{equation}
	\label{eqn:SZeq_barebasis}
	\mean{\sz}_\mathrm{sb} = \mean{\sz}_\mathrm{s} \frac{1}{\sqrt{1+4\lambda^2\lsb \mean{\ada}_s + \frac{\mean{\sz}_s+1}{2}\rsb}},	
\end{equation}
with $\mean{\sz}_\mathrm{s}$ given by Eq.~\eqref{eqn:mean_sz_s}.  The last term of Eq.~\eqref{eqn:szS} was neglected in the above expression as it oscillates rapidly in the rotating frame and  the operator $N_q$ has been replaced by its average value.  This expression for $\mean{\sz}_\mathrm{sb}$ is consistant with Eq.~(30) of Ref.~\cite{blais:2004a} when noticing that there $n$ corresponds to the total number of excitations while it is here the average number of photons.

It is interestingly to note that, because of the asymmetry of the expression for $\mean{\sz}_s$ with respect to $\ne$ and $\ng$, the steady-state value of $\sz$ depends on the measurement frequency.  Indeed, for a measurements of the phase where $\wm = \wr$, in the steady-state $n_{\e s} \approx n_{\g s} \approx \bar n_s$ and the effect will depend on the average number of photons in the resonator.  On the other hand, for amplitude measurements with $\wm = \wr + \chi$, $n_{\e s} \gg n_{\g s}$ and the departure from -1 should be less important.
% subsection measurement_induced_heat_bath (end)

\subsection{Comparison with exact numerics} % (fold)
\label{sub:comparison_with_exact_numerics}

To compare the results of the analytical expressions Eq.~\eqref{eqn:SZeq_barebasis} and \eqref{eqn:Gamma1Eff} to numerical integration of the full resonator-qubit master equation~\eqref{eqn:ExactMasterEquation}, we initialize the qubit in its excited state and the resonator in the corresponding steady-state with a continuous measurement drive of amplitude $\epm$ and frequency $\wr$.  In the absence of coherent driving at the qubit frequency, the qubit then simply decays to reach a steady-state value of $\sz$. By fitting the time evolution of $\mean{\sz}$ as obtained from numerical integration of \eq{eqn:ExactMasterEquation} to
\begin{equation}
	f(t) = A e^{-\gamma_{\rm eff} t} + \mean{\sz}_{sb},
\end{equation}
we extract the exact effective decay rate and steady-state mean value of $\sz$.

\begin{figure}
	\centering
	\includegraphics[width=0.95\hsize]{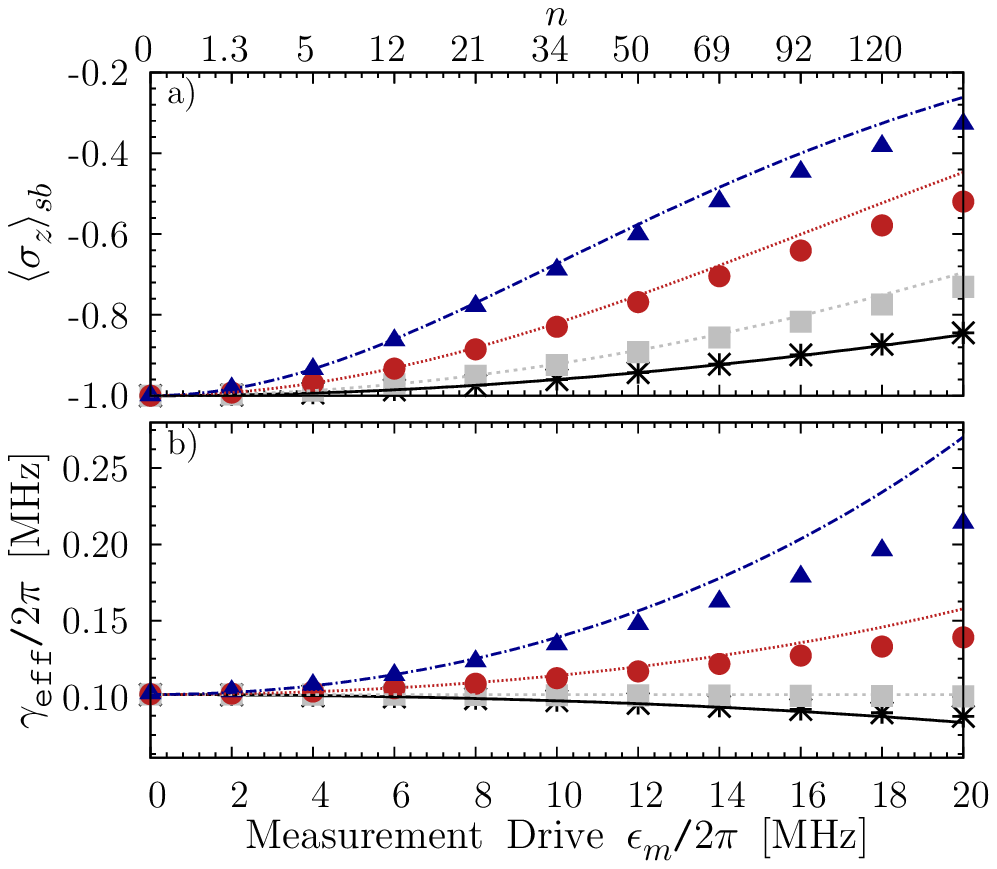}
	\caption{(Color online)  a) Steady state value of $\mean{\sz}$ and b) effective decay rate $\gamma_{\rm eff}$  as a function of measurement power for various values of $\gphi/2\pi = 0.0, 0.05, 0.2, 0.5$ MHz (black stars, grey squares, red circles, blue triangles). Symbols are extracted from numerical solution of Eq.~\eqref{eqn:ExactMasterEquation}.  Lines correspond to Eqs~\eqref{eqn:SZeq_barebasis} and \eqref{eqn:Gamma1Eff}. The numerical simulations were done with the same parameters as Fig.~\ref{fig:dtr_justify}, with the qubit initially in its excited state and the cavity in the corresponding steady-state with a continuous measurement drive of amplitude $\epm$. The top x-axis is the approximate average photons number in the cavity.}
	\label{fig:szeq_geff}
\end{figure}

These results are shown in Fig.~\ref{fig:szeq_geff},  the analytical expressions~\eqref{eqn:SZeq_barebasis} and \eqref{eqn:Gamma1Eff} (lines) in addition to the values extracted from numerical solution of the full master equation (symbols) are plotted.  With the parameters used here (see caption), the critical number of photons is $n_{\rm crit} = 400$ such that the figure shows results for $n/n_{\rm crit.} \lesssim 0.4$.

Figure~\ref{fig:szeq_geff}a) shows the steady-state value of $\mean{\sz}$ as a function of the measurement amplitude for increasing values of the pure dephasing rate $\gphi = \gamma_{\pm\Delta}/2$. The bottom line (black stars) corresponds to $\gphi = 0$, which in turns corresponds to the effective heat bath being at zero temperature.  As a result, in the dispersive basis, $\mean{\sz}_s = -1$ and the deviation from $-1$ is only caused by the change from dispersive to bare basis.  The lines lying above this result correspond to $\gphi/2\pi = 0.05, 0.2, 0.5$ MHz (dashed grey, dotted red, dashed-dotted blue).  Clearly, even for a relatively low number of photons compared to $n_{\rm crit.}$, mixing of the qubit excited and ground states by the effective heat bath can be significant if the pure dephasing rate is large.

While the reduced model is extremely accurate for $\gphi = 0$, it always slightly over-estimates $\mean{\sz}_\mathrm{sb}$ for non-zero dephasing rates.  Since the error is always positive, this can be interpreted as being due to the effect of a higher order terms in the dispersive approximation.  Nevertheless, the analytical model is at most  $\sim 10\%$ away from the exact numerical results for the range of parameters shown in Fig.~\ref{fig:szeq_geff}. 

Figure~\ref{fig:szeq_geff}b) shows the effective decay rate $\gamma_{\rm eff}$ as a function of measurement power for the same dephasing rates $\gphi$ as in pannel a).  The lines correspond to Eq.~\eqref{eqn:Gamma1Eff} while symbols are extracted numerically.  As expected from the discussion surrounding Eq.~\eqref{eqn:Gamma1Eff}, if the dephasing rate is negligible (black stars), the qubit effective decay rate falls below the bare decay rate $\gam/2\pi = 0.1$ MHz as the measurement power is increased.  However, for $2\gphi>\gam$ (red circles and blue triangles) the effective decay rate increases, again as expected from the model.  In this latter case, the photon number and dephasing dependent qubit mixing simply overwhelms the aforementioned decrease in $\gamma_{\rm eff}$ which is no longer visible.  For intermediate dephasing, $2\gphi = \gam$ (gray squares), these two processes cancel each other and the effective decay rate is almost independent of the measurement amplitude. 

%We note that the Purcell rate is not significant for the parameters used here, $\gamma_\kappa/2\pi \approx 1.5$kHz.  It only causes a small shift of the effective decay rate at zero measurement power.  Finally, we see that the effective model predicts an effective decay rate that is always further away from its bare value $\gam $ than seen numerically.  This again suggest that  higher order terms in the dispersive development tends to drag the decay rate toward its intrinsic value.
% subsection comparison_with_exact_numerics (end)
% section emergence_of_new_physics (end)

\section{Dispersive effects on the quantum trajectory equation} % (fold)
\label{sec:dispersive_effects_on_the_stochastic_master_equation}
The master equation description of the dynamics does not take into account the result of the measurement.  To include this information, we use quantum trajectory theory~\cite{gardiner:2004b,carmichael:1993a,wiseman:1993a,gambetta:2005a} and derive the evolution equation for the conditional state, or quantum trajectory equation (QTE).  This was already study in Ref.~\cite{gambetta:2008a} for the linear dispersive model and is extended here to incorporate the non-linear effects. 

When monitoring the resonator bath, characterized by the rate $\kappa$, an observer would in principle see two signals: one at the qubit frequency and one at the cavity frequency \footnote{When monitoring the qubit baths there would also be two signals for both the $\gamma_1$ and $\gamma_\phi$ baths}.  To derive the QTE, it is assumed as above that the relevant bath  frequencies for the resonator bath are well separated such that they can be treated as two separate Markovian baths with relevant frequencies $\omega_r$ and $\omega_a$.  That is, for an infinitesimal interval $dt$ the unitary operator describing the resonator bath is given by Eq.~\eqref{eq:Unitkappa}.  

In a homodyne measurement, with a local oscillator set to $\sim\omega_r$,  the bath is projected in an eigenstate of the operator 
$dB_{\kappa,r} +  dB^{\dg}_{\kappa,r}$ (where $dB_{\kappa,r}$ is defined in appendix \ref{annsec:obtaining_the_dispersive_master_equation}), with measurement result $J$~\cite{gardiner:2004b}.  For many such measurements, each separated by a time $dt$ and with result $J_k$, the state conditioned on the complete record $\mathbf{J}(t) = \{J_1,..., J_k\}$ can be expressed as~\cite{gambetta:2005a}
\begin{equation}
	\varrho_\mathbf{J}(t) = \tilde\varrho_\mathbf{J}(t)/\mathrm{Pr}(\mathbf{J}),
\end{equation} 
where $\mathrm{Pr}(\mathbf{J}) $ is the probability for observing record \textbf{J}
\begin{equation}
	\mathrm{Pr}(\mathbf{J}) = \mathrm{Tr}[\tilde\varrho_\mathbf{J}(t)] 
\end{equation}
and $\tilde\varrho_\mathbf{J}(t)$ is an unnormalized conditional state given by
\begin{equation}\label{eq:stepqt}
	\begin{split}
		\tilde\varrho_\mathbf{J}(t) =& \sum_{\mathbf{F}}M_{J_k,\mathbf{F_k}}(dt)...M_{J_1,\mathbf{F_1}}(dt)\varrho(0)M\dg_{J_1,\mathbf{F_1}}(dt)\\&...M\dg_{J_k,\mathbf{F_k}}(dt).
	\end{split}
\end{equation}
Following the notation of Ref.~\cite{gambetta:2005a}, $M_{J_k,\mathbf{F_k}}$ is the Kraus operator of the $k^\text{th}$ measurement and $\mathbf{F_k} = \{f_{2k},f_{3k},f_{4k}\}$ the results of fictitious measurements performed on the baths coupled to the operators $L_2$, $L_3$ and $L_4$ defined in  \eq{eq:bathrwa}.  Since these latter measurements are fictitious and the conditional state does not depend on them (we sum over all fictitious results to obtain $\tilde\varrho_\mathbf{J}$), it is possible to choose the fictitious observable at will~\cite{gambetta:2005a}.  For simplicity, we assume fictitious homodyne measurement of the unobserved operators $L_j$ with $j=2,3,4$.  Using the evolution operator \eq{eq:Utotal}, this leads to the following Kraus operator
\begin{equation}\label{eq:measop}
	\begin{split}
			M_{J,\mathbf{F}}(dt) =& \bra{\mathbf{F},J} U(t+dt,t)\ket{0,0}\\
			=&\sqrt{\Upsilon_{J,\mathbf{F}}} 
			[1-i H_\mathrm{tot} dt 
			+ \sqrt{\kappa} L_1 Jdt + \kappa  L_1^\dagger L_1  dt/2\\
			&+\mathbf{F}^T\mathbf{L}dt-\mathbf{L}^\dag\mathbf{L}dt/2]
			%+ \sqrt{\kappa} L_1 Jdt \\
			%& + \sqrt{\gamma_\kappa}\sigma_- f dt- \kappa  L_1^\dagger L_1  dt-\gamma_\kappa \sigma_+\sigma_-dt]
	\end{split}
\end{equation}
where $\mathbf{L} = \{\sqrt{\kappa_a} L_2, \sqrt{\gamma_a} L_3, \sqrt{\gamma_r}L_4\}$ and $\Upsilon_{J,\mathbf{F}}$ is the gaussian probability measure
\begin{equation}
	\Upsilon_{J,\mathbf{F}} dJdf = \frac{1}{(2\pi/dt)^2}\exp[-(J^2+\mathbf{F}^2)dt/2]dJd\mathbf{F}.
\end{equation}

For continuous monitoring, the time step $dt$ between measurements tend towards 0. In this limit, Eq.~\eqref{eq:stepqt} leads to the QTE whose ensemble average is the unconditional master equation. The corresponding QTE, in \ito form, for the measurement operator Eq. \eqref{eq:measop} is 
	\begin{equation}
		\label{eqn:ConditionalMasterEquationFull}
		\begin{split}
			\dot \crho_J\trans{\atD} =& \sL\trans{\atD}\crho_J\trans{\atD} + 2 \sqrt{\kappa\eta}{\cal M}[ I_\phi(1+\lambda^2\sz/2)]\crho_J\trans{\atD}\xi(t) \\							
			&+i\sqrt{\kappa\eta}[Q_\phi(1+\lambda^2\sz/2),\crho_J\trans{\atD}] \xi(t)
		\end{split}
	\end{equation}	
where $\sL\trans{\atD}$ is given by Eq.~\eqref{eqn:ExactMasterEquation_D}.	In this expression, we have defined the $\phi$-dependent field quadratures $2I_\phi = a e^{-i\phi} + a \dg e^{i\phi}$ and $2Q_\phi = -iae^{-i\phi} + i\ad e^{i\phi}$. Moreover, the superoperator ${\cal M}[c]$ is defined as 
\begin{equation}
		{\cal M} [c]\crho= (c -\mean{c}_t) \crho/2+\crho(c-\mean{c}_t)/2,
\end{equation}
where $\mean{c}_t = \trace{c\crho_J\trans{\atD}(t)}$ and the measurement outcome, $J$ can be expressed as
\begin{equation}
	J(t) =2 \sqrt{\kappa\eta}\mean{I_\phi (1+\lambda^2\sz/2)}_t + \xi(t),
\end{equation}
where $\xi(t)$ is Gaussian white noise and $\eta$ is a detection efficiency parameter included for completeness~\cite{gambetta:2008a}.

While only the signal at the cavity frequency was taken into account here, it is interesting to point out that the signal at the qubit frequency could also be measured.  This was done experimentally in Ref.~\cite{houck:2007a} to perform qubit state tomography.  However, the signal at that frequency is in general much weaker than the signal at the cavity frequency.  As a result, while by itself the former signal would lead to a very inefficient QTE (which would have a similar to that of a  direct homodyne measurement of the qubit) this additional information could be included in the present treatment to realize even more efficient qubit measurements.

%Note that in the above we have considered conditioning on the signal at the cavity frequency, in fact we could consider both the signal at the cavity frequency and the qubit frequency or just the qubit frequency. The reason why we only consider the signal at the cavity frequency is that the signal at the qubit frequency will be much smaller then that at the cavity (approximately $g/\Delta$ due to cavity filtering) and as a result if we only conditioned on this frequency we would get a very inefficient SME, by this we mean the conditional state would not vary much from the solution of the master equation. However there is still information there as it was this information that what was measured in the Ref. \cite{houck:2007a} to do state tomography of the qubit. Thus if we were to use this information we would get an extra stochastic term in the Eq. \eqref{eqn:ConditionalMasterEquationFull} which would have a form similar to that found in quantum optics for monitoring the fluorescence of a two level atom (see Refs. \cite{carmichael:1993a,gambetta:2001a}) and one could do a better measurement of the state of the qubit.  

\subsection{Effective qubit quantum trajectory equation} % (fold)
\label{sub:the_effective_qubit_stochastic_master_equation}
Using the polaron transformation~\eq{eqn:PolaronTransformation}, it is possible to obtain a reduced QTE for the qubit.  As show in appendix~\ref{sec:derivation_of_the_effective_qubit_quantum_trajectory}, this reduced QTE takes the form
\begin{equation}
	\label{eqn:conditional_qubit_master_equation}
	\begin{split}
		\dot\qrho_{\bar J}\trans{\atD} &= \sL\trans{\atD} \qrho_{\bar J}\trans{\atD} + \sqrt{\Gamma_\mathrm{ci}(t)} \sM[\sz]\qrho_{\bar J}\trans{\atD}(t) (\bar J(t) - \sqrt{\Gamma_\mathrm{ci}(t)}\mean{\sz}_t) \\
		&\quad -i\frac{\sqrt{\Gamma_\mathrm{ba}(t)}}{2} \com{\sz}{\qrho_{\bar J}\trans{\atD}(t)}(\bar J(t) - \sqrt{\Gamma_\mathrm{ci}(t)}\mean{\sz}_t),
	\end{split}
\end{equation} 
where  $\Gamma_\mathrm{ci}(t)$ is the rate at which information comes out of the resonator and $\Gamma_\mathrm{ba}(t)$ represents extra non-Heisenberg back-action from the measurement.  $\bar J(t)$ is the processed record coming from the resonator and is given by~\cite{gambetta:2008a}
\begin{equation}
	\label{eqn:measurement_record}
	\bar J(t) = \sqrt{\Gamma_{ci}} \mean{\sz}_t + \xi(t).
\end{equation}
This quantity is linked to the homodyne current by
\begin{equation}
	\label{eqn:homodyne_current}
	\begin{split}
		J(t) = \bar J(t) &+ \sqrt{\kappa\eta} \lvb\mu(t)\rvb\cos(\theta_\mu-\phi)\\
		&+\sqrt{\kappa\eta} \frac{\lambda^2}{2} \lvb\beta(t)\rvb\cos(\theta_\beta-\phi).
	\end{split}
\end{equation}
Eq.~\eqref{eqn:conditional_qubit_master_equation} has the same form as the QTE found in Ref.~\cite{gambetta:2008a} for the linear model, appart from second order corrections to the rates $\Gamma_\mathrm{ci}(t)=\eta\Gamma_m\cos^2(\theta_m)$ and $\Gamma_\mathrm{ba}(t)=\eta\Gamma_m\sin^2(\theta_m)$, where
\begin{subequations}
	\begin{align}
		\Gamma_m &= \kappa\lvb\beta\rvb^2\lp 1+\frac{\lvb\mu\rvb\cos(\theta_\beta-\theta_\mu)}{4\lvb\beta\rvb n_\mathrm{crit.}} + \frac{\lvb\mu\rvb^2}{64\lvb\beta\rvb^2 n_\mathrm{crit.}^2}\rp, \\
		\theta_m &= \phi-\theta_\beta + \ImaginaryPart \lcb \ln\lsb 1+\frac{\lvb\mu\rvb e^{i(\theta_\beta-\theta_\mu)}}{8\lvb\beta\rvb n_\mathrm{crit.}}\rsb\rcb.
	\end{align}
\end{subequations}
These expressions are valid for $\phi-\theta_\beta \elem\lsb 0,\pi/2\rsb$, with $\theta_\beta = \arg(\beta)$, $\theta_\mu = \arg(\mu)$ and $\beta$, $\mu$ defined in \eq{eqn:beta_mu}. The second order corrections do not change the physics in an important way since, as in the linear model, it is possible to choose the phase $\phi$ of the LO  optimally such that $\Gamma_\mathrm{ba}(t)$ is zero.  The corrections have the effect of reducing $\Gamma_\mathrm{ci}(t)$ in comparison to what is obtained in the linear model~\cite{gambetta:2008a}. 

To demonstrate the different features of the reduced QTE, Fig.~\ref{fig:heat_jumps} presents
\begin{figure}
	\centering
	\includegraphics[width=0.95\hsize]{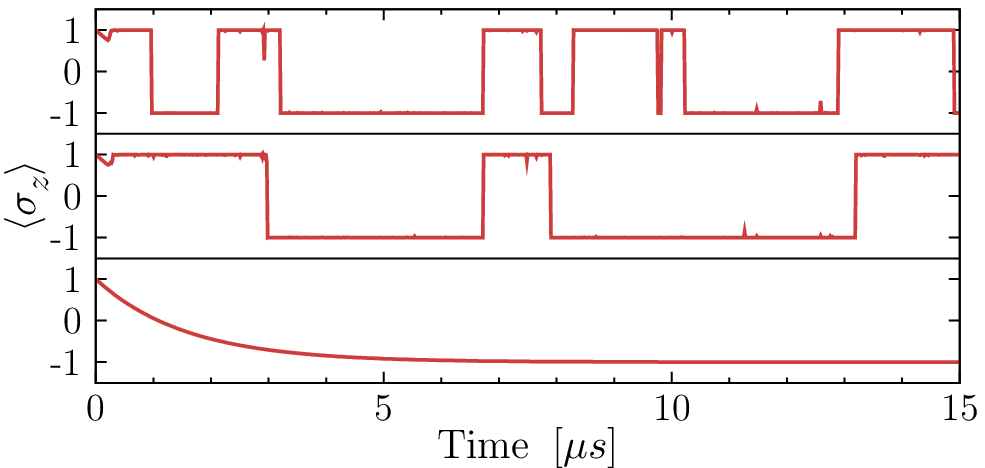}
	\caption{Typical trajectories for $\gphi/2\pi=0.5$MHz and measurement amplitudes of $\epm/2\pi=0$ MHz (bottom), $\epm/2\pi=10$ MHz (center) and $\epm/2\pi=20$ MHz (top).  The other parameters are the same as in Fig.~\ref{fig:dtr_justify}. The initial state has $\mean{\sz}=1$ with zero photons and a measurement drive starting shortly after $t=0$.}
	\label{fig:heat_jumps}
\end{figure}
a typical trajectory for 3 different measurement powers using the same parameters as in Fig.~\ref{fig:dtr_justify}.  As in the linear case, increasing the measurement power localizes the qubit state  on one of its basis states.   As a result, although the QTE is based on homodyne measurement, we do not expect diffusive but rather jump-like trajectories.  Moreover, because of the effective upward rate $\gu$ which increases with measurement power,  the trajectories show telegraph noise rather than a single jump to the ground state. These predictions can be experimentally tested once single shot measurement is achievable.  From these results, the waiting time between jumps can be compared to $\gu$ and $\gd$.

The SNR can be defined as
\begin{equation}
	\mathrm{SNR}=\frac{\Gamma_\mathrm{ci}}{\gamma_\mathrm{eff}}
\end{equation} 
which at the optimal point is $\eta\Gamma_\mathrm{m}/\gamma_\mathrm{eff}$. As in the linear model, $\Gamma_\mathrm{m}$ is proportional, through $\lvb\beta\rvb^2$, to the cavity pull $\chi+\zeta\ada$ times the number of photons in the cavity. However, and contrary to the linear model, the cavity pull {\em decreases} with increasing photon population. Moreover, we have shown that the measurement photons act as a heat bath with $\gamma_\mathrm{eff}\propto\bar n$. Therefore, unlike the linear predictions, which predicts the SNR to increase linearly with photon population, we expect the SNR to saturate at higher photon numbers. This is discussed in more detail in Ref.~\cite{boissonneault:2008a}.  It is important to point out that the main contribution to this effect is the reduction of the cavity pull, and not the measurement-enhanced mixing rate.

% subsection the_effective_qubit_stochastic_master_equation (end)

% section dispersive_effects_on_the_stochastic_master_equation (end)

\section{Conclusion} % (fold)
\label{sec:conclusion}

We have investigated circuit QED in the dispersive regime.  To take into account large photon population of the resonator, useful for qubit readout, we have shown that it is necessary to push the dispersive treatment to a higher order.  We have done this while taking into account the effect of dissipation and external microwave driving.  In particular, we have obtained a Markovian model for the effect of dissipation that takes into account frequency dependence of the environment.

Building on our previous work~\cite{gambetta:2008a}, we have then traced over the resonator states to obtain an effective master equation for the qubit  valid in the limit $n\ll \left\{n_\kappa=\kappa/\zeta, n_\mathrm{crit} = \Delta^2/4g^2 \right\}$. A striking feature of the resulting master equation is that the qubit relaxation and dephasing rates now depend on the number of photons populating the resonator.   Moreover, in the presence of pure dephasing, we have shown that measurement will cause excitation of the qubit.  In other words, the photon population of the resonator act as an effective heat bath on the qubit.  This can lower the effective `quality' of the dispersive QND readout of the qubit.  

Finally, using the quantum trajectory approach, we have obtained an effective stochastic master equation for the qubit. In the single-shot limit, this equation predicts that the measurement-induced heat bath should lead to telegraph-like jumps in the measurement response.  Moreover, the non-linearity have been shown to lead to a reduction of the expected signal-to-noise ratio, a result qualitatively consistent with experimental observations~\cite{boissonneault:2008a}.

There are various ways to test experimentally these predictions. First, the qubit effective decay rate $\gu(n)+\gd(n)$ and the steady-state value of $\mean{\sz}$ can be measured experimentally and compared to the results obtained here.  Second, when single shot measurements become possible in circuit QED, the effect of the upward transitions caused by the measurement induced heat bath should be observed.  The waiting times between the upward and downward transitions can then be related to the rates $\gu(n)$ and $\gd(n)$ obtained here.

% section conclusion (end)

\begin{acknowledgments}
We thank J.~Koch, A.~A.~Houck, R.~J.~Schoelkopf and S.~M.~Girvin for discussions.
MB was supported by the Natural Sciences and Engineering Research Council of Canada (NSERC).
AB was supported by the Natural Sciences and Engineering Research Council of Canada (NSERC), the Fond Qu\'eb\'ecois de la Recherche sur la Nature et les Technologies (FQRNT) and the Canadian Institute for Advanced Research (CIFAR).  
JMG was supported by CIFAR, MITACS, and ORDCF.
%, the NSA and LPS under ARO W911NF-05-1-0365, the NSF under DMR-0653377, Australian Research Council (FF0458313 and CE0348250), and the State of Queensland.
 
\end{acknowledgments}

%%%%%%%%%%%%
\appendix
%%%%%%%%%%%%

\section{Exact diagonalisation of the Jaynes-Cummings Hamiltonian by unitary transformation} % (fold)
\label{sct:ExactDiag}
In this appendix, the Jaynes-Cummings Hamiltonian is diagonalized exactly using a unitary transformation. In order to simplify the notation, we introduce the commutation linear application
\begin{equation}
	\sC_A B \equiv \com{A}{B}, \quad \sC_A^m B = \stackrel{m\ times}{[A,[A,[A\dots},B]]].
\end{equation}
In term of this superoperator, Hausdorff's relation can be written as
\begin{equation}
	e^A B e^{-A} = \sum_{n=0}^{\inf} \frac{1}{n!} \sC_A^n B.
\end{equation}
In the same way as for the linear approximation discussed in section~\ref{sub:dispersive_jaynes_cummings_hamiltonian_linear_regime}, the antihermitian operator $\im$ will be key in this diagonalization.  Since it commutes with both $H_s$ and $\im$, another important operator is the total number of quanta  $N_q$ defined in Eq.~\eqref{eqn:N_q}.  Moreover, with a unitary operator of the form 
\begin{equation}
	\tD = e^{-\Lambda(N_q) \im},
\end{equation}
where $\Lambda$ is a function to be defined, $N_q$ can be considered as a scalar when applied on $H_s$. 

Before transforming $H_s$ using $\tD$, it is useful to introduce some important commutators. First, it is simple to show that
\begin{equation}
	\sC_\im H_0 = \hbar\Delta \ip.
\end{equation}
Using this result, transformation of $H_s$ by $\tD$ yields
\begin{equation}
	H_s\trans{\atD} \equiv \tD^\dag H_s\tD = H_0 + \hbar\sum_{n=0}^\inf \frac{(n+1)g + \Delta\Lambda}{(n+1)!} \sC_{\Lambda\im}^n \ip.
\end{equation}
with
\begin{align}
	\sC_{\Lambda\im}^{2n}\ip &= (-4)^n \Lambda^{2n} N_q^n \ip \\
	\sC_{\Lambda\im}^{2n+1}\ip &= -2 (-4)^n \Lambda^{2n+1} N_q^{n+1} \sz.
\end{align}
Using this last result, we find
\begin{equation}
	\begin{split}
	 H_s\trans{\atD} &= H_0 + \hbar\lp \frac{\Delta \sin{\lp 2 \Lambda \sqrt{N_q}\rp}}{2\sqrt{N_q}} + g\cos{\lp 2 \Lambda \sqrt{N_q}\rp} \rp \ip \\
	&- 2\hbar N_q\sz \lp \frac{g \sin\lp 2 \Lambda \sqrt{N_q}\rp}{2\sqrt{N_q}} + \frac{\Delta\lp1- \cos\lp2 \Lambda \sqrt{N_q}\rp\rp}{4N_q} \rp.		
	\end{split}
\end{equation}
To complete the diagonalization, we take
\begin{equation}
	\Lambda(N_q) = \frac{-\arctan\lp2\lambda\sqrt{N_q}\rp}{2\sqrt{N_q}}.
\end{equation}
such as to eliminate the off-diagonal term proportional to $\ip$.  Using this result, we finally obtain the exact diagonal form
\begin{equation}
	H_s\trans{\atD} = H_0 - \frac{\hbar\Delta}{2} \lp 1 - \sqrt{1+4\lambda^2N_q} \rp \sz.
\end{equation}

Using this result, we define the Lamb and ac-Stark shift operators as [we use $H_s\trans{\atD}(\ada,\sz)$]
\begin{align}
	\begin{split}
		\label{eqn:LS}
		\raisetag{32pt}
		\delta_L &\equiv H_s\trans{\atD}(0,1) - H_s\trans{\atD}(0,-1) - \hbar\wa \\
		& = - \frac{\hbar\Delta}{2} \lp 1-\sqrt{1+4\lambda^2}\rp
	\end{split}\\
	\begin{split}
		\raisetag{16pt}
		\label{eqn:SS}
		\delta_S(\ada) &\equiv H_s\trans{\atD}(\ada,1) - H_s\trans{\atD}(\ada,-1) - \delta_L - \hbar\wa \\
		&= \frac{\hbar\Delta}{2} \lp \sqrt{1+4\lambda^2(\ada+1)} + \sqrt{1+4\lambda^2\ada} \right. \\
		& \qquad\qquad \left. - 1 - \sqrt{1+4\lambda^2} \rp.
	\end{split}
\end{align}
Developing these expressions in powers of $\lambda$, we obtain
\begin{align}
	\label{eqn:LambShift_approx}
	\delta_L &\approx \hbar \chi + \mathcal{O}(\lambda^5) \\
	\label{eqn:StarkShift_approx}
	\delta_S(\ada) &\approx \hbar \chi \ada + \hbar\zeta (\ada)^2 + \mathcal{O}(\lambda^5),
\end{align}
with $\chi = g\lambda(1-\lambda^2)$ and $\zeta = -g^4/\Delta^3$.  These approximate results are used in  Eqs.~\eqref{eqn:ApproxHDisp} and \eqref{eqn:H_s_D}.
% section exact_diagonalisation_of_the_jaynes_cummings_hamiltonian_by_unitary_transformation (end)

\section{Obtaining the dispersive master equation} % (fold)
\label{annsec:obtaining_the_dispersive_master_equation}

\subsection{Qubit relaxation and photon decay} % (fold)
\label{annsub:qubit_relaxation_and_photon_decay}
In this appendix, we find the effect of the dispersive transformation on the non-unitary part of the master equation. Applying the dispersive transformation on the Hamiltonians Eqs.~\eqref{eq:bathcoupling}, moving to the interaction frame defined by the transformation $\exp[-i (H_s+H_{B\kappa}+H_{B\gamma})t/\hbar]$ and performing a rotating-wave approximation (RWA) yields
\begin{equation}
	\label{eq:bathrwa}
	\begin{split}
		H_\kappa\trans{\atD} &= i\hbar \lsb L_1 z^\dagger_{\kappa}(t,\omega_r) + L_2 z^\dagger_{\kappa}(t,\omega_a)\rsb +\mathrm{h.c.} \\
		H_\gamma\trans{\atD} &= i\hbar \lsb L_3 z^\dagger_{\gamma}(t,\omega_a) + L_4 z^\dagger_{\gamma}(t,\omega_r) \rsb +\mathrm{h.c.},
	\end{split}
\end{equation}
with $L_1 = a(1+\lambda^2\sigma_z/2)$, $L_2 = \lambda\sm$, $L_3= \sm [1-\lambda^2(\ada + 1/2)]$ and $L_4= \lambda a\sz$. The bath operators $z_{i}(t,\omega_p)$ are given by
\begin{equation}
	z_{i}(t,\omega_p) = \int_{\omega_p - B_{i,p}}^{\omega_p + B_{i,p}} \sqrt{d_i(\omega)} f_i(\omega)b_i(\omega) e^{-i(\omega - \omega_p )t} d\omega.
\end{equation}
We have kept in Eq.~\eqref{eq:bathrwa} only terms that will contribute up to order $\lambda^2$ in the master equation. In obtaining this expression, we have taken the (dispersive) system Hamiltonian as $H_s \approx \hbar\omega_r\ada + \hbar\omega_a\sz/2$, where $\omega_a$ should be understood as the Lamb and ac-Stark shifted qubit transition frequency and $\omega_r$ should be understood as the cavity frequency shifted by the non-linearity (i.e. $\omega_r+\zeta$).  Moreover, to perform the RWA, we have made the standard assumption that the system-bath interaction is limited to a small band of frequency $B_{i,p}$ around the frequency $\omega_p$ of the corresponding system operator $L_i$, $B_{i,p}\ll\omega_p$~\cite{gardiner:2004b}. %We also note that we have omitted the effect of control drives on the qubit. This is reasonable if the control drive duration is short. The effect of continuous drive on the decay rates was studied, for example, in Ref~\cite{ithier:2005}.
% AB: I have removed this completely since we don't deal with control at all in this paper.

We now assume that, within the bandwidths $B_{i,p}$, the coupling constants $f_i(\omega_p)$ and the density of modes $d_i(\omega)$ do not vary significantly.  In this situation, the above system-bath Hamiltonians can be rewritten as 
\begin{equation}
	\begin{split}
		H_\kappa\trans{\atD} &= i\hbar \sqrt{\kappa_r} L_1 b^\dag_{\kappa,r}(t) +i\hbar \sqrt{\kappa_a}L_2 b^\dag_{\kappa,a}(t) + \mathrm{h.c.} \\
		H_\gamma\trans{\atD} &= i\hbar \sqrt{\gamma_a} L_3 b^\dag_{\gamma,a}(t) +
		i\hbar \sqrt{\gamma_r} L_4 b^\dag_{\gamma,r}(t) + \mathrm{h.c.} 
	\end{split}
\end{equation}
where the decay rates are given by~\eq{eqn:rates} and the bath temporal modes are defined as
\begin{equation}
	\label{eq:temporal}
	b_{i,p}(t) = \frac{1}{\sqrt{2\pi}}\int_{\omega_p-B_{i,p}}^{\omega_p+B_{i,p}} d\omega b_i(\omega)e^{-i(\omega-\omega_p)t}.
\end{equation}

Since $\com{b_i(\omega)}{b^\dag_j(\omega')} = \delta_{i,j}\delta(\omega-\omega')$, the commutator of two temporal modes is (after a change of integration variable)
\begin{equation}\label{eq:corr}
	\begin{split}
	&[b_{i,p}(t),b_{j,q}^\dag(t')] \\
	 &= \frac{\delta_{i,j}}{2\pi}\int_{-B_{i,p}}^{ B_{i,p}}\int_{-B_{i,q}}^{ B_{i,q}} 
	d\omega d\omega'  \delta(\omega-\omega'-\omega_p+\omega_q)
	e^{-i\omega t}e^{-i\omega' t'}.
	\end{split}
\end{equation}
If we now take $|\omega_p-\omega_q|\gg B_{i,p},B_{i,q}$ for $p\neq q$ then the above becomes
\begin{equation}
	[b_{i,p}(t),b_{j,q}^\dag(t')] \\
	= \frac{\delta_{i,j}\delta_{p,q}}{2\pi}\int_{-B_{i,p}}^{ B_{i,p}}
	d\omega	e^{-i\omega (t-t')}.
\end{equation}
In other words, we assume the bath operators to be independent.  In the dispersive regime, this is a reasonable assumption since $|\omega_p-\omega_q|\sim \Delta$, where the detuning $|\Delta|\gg g$ is large. % (see Fig.~\ref{fig:Spectrums} a).

We finally make the standard and reasonable assumption that dissipation is not too strong, such that the time scales set by the decay rates $\kappa_p$ and $\gamma_p$ are much longer than the cutoff time $1/B_{i,p}$.  In this situation we can effectively take the limit $B_{i,p}\rightarrow\infty$.  This corresponds to the standard Markov approximation~\cite{gardiner:2004b}, which was already successfully applied to describe circuit QED experiments~\cite{wallraff:2004a,schuster:2005a,wallraff:2005a,schuster:2007a,houck:2007a,majer:2007a}. In this situation, the above commutation relation reduces to
\begin{equation}\label{eq:corr2}
	[b_{i,p}(t),b_{j,q}^\dag(t')] \\
	= \delta_{i,j}\delta_{p,q}\delta(t-t').
\end{equation}

In this Markov, or white noise, approximation the evolution operators can be written in It\^o form as
\begin{equation}\label{eq:Utotal}
	U(t+dt,t) = U_\kappa(t+dt,t) U_\gamma(t+dt,t) e^{-i H_\mathrm{tot} dt}
\end{equation}
with
\begin{subequations}
	\begin{align}
		\label{eq:Unitkappa}
		\begin{split}
			U_\kappa(t+dt,t) &= \exp\left\{-i\sqrt{\kappa_r}[L_1 dB^\dag_{\kappa,r}-L_1^\dag dB_{\kappa,r}]\right.\\
			&\quad \left.-i\sqrt{\kappa_a}[L_2 dB^\dag_{\kappa,a}-L_2^\dag dB_{\kappa,a}]\right\}U_\kappa(t)			
		\end{split} \\
		\begin{split}
			U_\gamma(t+dt,t) &= \exp\left\{-i\sqrt{\gamma_a}[L_3 dB^\dag_{\gamma,a}-L_3^\dag dB_{\gamma,a}]\right.\\
			&\quad\left.-i\sqrt{\gamma_r}\lambda[L_4 dB^\dag_{\gamma,r}-L_4^\dag dB_{\gamma,r}]\right\}U_\gamma(t),			
		\end{split}
	\end{align}
\end{subequations}
where $dB_{i,p} = b_{i,p}dt$ is a quantum Wiener increment~\cite{gardiner:2004b}. 

We now take the bath to be in the vacuum state and uncorrelated to the system at time $t=0$.  By tracing over the bath and keeping terms of order $\mathcal{O}(dt)$ using \ito calculus, we obtain a Lindblad form master equation for the resonator-qubit system.  In this master equation, the photon bath $\kappa$ leads to the damping superoperators~\cite{gardiner:2004b}
\begin{equation}
	\label{eq:master_photon_bath}
	\kappa\sD[a(1+\lambda^2\sz/2)]\crho\trans{\atD} + \gamma_\kappa\sD[\sm]\crho\trans{\atD}
\end{equation} 
while the qubit bath $\gamma$ leads to
\begin{equation}
	\label{eq:master_qubit_bath}
	\gamma\sD\left[\sm\{1-\lambda^2(\ada+1/2)\}\right]\crho\trans{\atD} + \kappa_\gamma\sD[a\sz]\crho\trans{\atD}.
\end{equation}
 These terms are the second and third lines of Eq.~\eqref{eqn:ExactMasterEquation_D}.
% subsection qubit_relaxation_and_photon_decay (end)

\subsection{Qubit dephasing} % (fold)
\label{annsub:qubit_dephasing}

For dephasing, we start with the Hamiltonian~\eqref{eq:classical_dephasing_omega}. Moving to the dispersive basis, it becomes
\begin{equation}
	H\trans{\atD}_\mathrm{dep} 
	= \hbar\nu \left[ \sz (1 - 2 \lambda^2 N_q )- 2 \lambda \ip \right] 
	\int_{-\infty}^\infty f_\varphi(\omega) e^{i\omega t}d\omega,
\end{equation} 
where we have used the second order expansion of \eq{eqn:szS}. Moving to a frame rotating at the qubit and resonator frequencies we find
\begin{equation}
	\begin{split}
		H_\mathrm{dep}\trans{\atD} &= \hbar\nu\sz (1 - 2 \lambda^2 N_q ) f_0(t)- 2\hbar\nu \lambda \ad\sm
		 f_{\Delta}(t)\\
		&\quad -2\hbar\nu\lambda a\sp f_{-\Delta}(t),
	\end{split}
\end{equation}
where
\begin{equation}
	f_{\omega_0}(t) = \int_{-\infty}^{\infty} f_\varphi(\omega) e^{i(\omega-\omega_0) t}d\omega.
\end{equation}
The main contribution to dephasing comes from a small frequency band $B_0$ centered around the frequency $\omega_0$.  In this situation, the integration boundaries in $f_{\omega_0}(t)$ can be reduced to
\begin{equation}\label{eq:noise}
	f_{\omega_0}(t) = \int_{\omega_0-B_0}^{\omega_0+B_0} f_\varphi(\omega) e^{i(\omega-\omega_0) t}d\omega.
\end{equation}
For this rotating-wave approximation to be valid, it is required that $B_0 \ll \omega_0$~\cite{gardiner:2004b}.

The Wiener-Khinchin theorem can be used to relate $f_\varphi(\omega)$ to its noise spectrum $S(\omega)$~\cite{gardiner:2004c}
\begin{equation}
	E[f_\varphi(\omega)f_\varphi(-\omega')] = \delta(\omega-\omega')S(\omega),
\end{equation}
where $E[\cdot]$ is an ensemble average.  This allows us to write the $\omega$ component of the noise as
\begin{equation}
	f_\varphi(\omega) = \sqrt{S(\omega)}\xi(\omega),	
\end{equation}
with $\xi(\omega)$ white noise obeying $E[\xi(\omega)] = 0$ and $E[\xi(\omega)\xi(-\omega')] = \delta(\omega-\omega')$. 

Using these results, we now make similar assumptions as in the last section and take the noise spectrum $S(\omega)$ to be constant within the small frequency band $B_0$ around $\omega_0$.  After a change of integration variable, $f_{\omega_0}(t)$ can be written as
\begin{equation}
	f_{\omega_0}(t) = \sqrt{S(\omega_0)} \int_{-B_0}^{B_0} \xi(\omega+\omega_0) e^{i\omega t}d\omega.
\end{equation}
In this Markov approximation, we will again assume the noise spectrum to be relatively weak which implies that the time scale corresponding to dissipation is much slower than $1/B_0$~\cite{gardiner:2004b}.  In this situation, we take $B_0\rightarrow\infty$ which allows us to write
\begin{equation}
	f_{\omega_0}(t) =\sqrt{S(\omega_0)}\xi_{\omega_0}(t)
\end{equation}
such that the transformed dephasing Hamiltonian becomes 
\begin{equation}
	\begin{split}
	H_\mathrm{dep}\trans{\atD} 
	&= \hbar\nu\sqrt{S(0)}\sz (1 - 2 \lambda^2 N_q ) \xi_0(t) \\
	&\quad - 2\hbar\nu\sqrt{S(\Delta)} \lambda \ad\sm \xi_{\Delta}(t) \\
	&\quad -2\hbar\nu\sqrt{S(-\Delta)}\lambda a\sp\xi_{-\Delta}(t).
	\end{split}
\end{equation}
The three $\xi_{\omega_0}(t)$ white noise terms in the above expression now correspond to independent noises, centered around three different frequencies.  

The above Hamiltonian leads to the following superoperators in the resonator-qubit master equation
\begin{equation}
	\label{eqn:master_dephasing_bath}
	\begin{split}
			&\gphi\sD[\sz \{1 - 2 \lambda^2 (\ada+1/2) \}]\crho\trans{\atD}/2 \\
			&+ \gamma_\Delta \sD[\ad\sm]\crho\trans{\atD} + \gamma_{-\Delta} \sD[a\sp]\crho\trans{\atD},
	\end{split}
\end{equation} 
with the rates given by Eq.~\eqref{eqn:rates}. These terms correspond to the fourth and fifth lines of Eq.~\eqref{eqn:ExactMasterEquation_D}.
% subsection qubit_dephasing (end)

% section obtaining_the_dispersive_master_equation (end)

\section{The polaron transformation} % (fold)
\label{sec:the_polaron_transformation}
Following the approach developed in Ref~\cite{gambetta:2008a}, a reduced master equation for the qubit is obtained in this appendix. To do so, we start from the dispersive master equation~\eqref{eqn:ExactMasterEquation_D} and go to the rotating frame defined by $\tR = \exp[i\wm \ada t]$. We then go to a frame defined by the polaron-type transformation
\begin{equation}
	\label{eqn:PolaronTransformation}
	\tP = \Pe D(\ae) + \Pg D(\ag),
\end{equation}
where $D(\alpha)$ is the displacement operator and $\alpha_{g(e)}$ satisfy Eq.~\eqref{eqn:Condition_alphas}. In the polaron frame, the field $a$ is described by a classical part given by the complex variables $\alpha_g$ and $\alpha_e$, and a small quantum part corresponding to quantum noise.  

The action of $\tP$ on various system operators is given by
\begin{subequations}
	\begin{align}
		\tP^\dag a \tP &= a + \Pa \\
		\tP^\dag \ada \tP &= \ada + \ad\Pa + a\Pa^* + \lvb\Pa\rvb^2 \\
		\tP^\dag \sm \tP &= \sm \Dop^\dag(\ag)\Dop(\ae) \\
		\tP^\dag \sz \tP &= \sz \\
		\label{eqn:adaada_P}
		\begin{split}
			\tP^\dag(\ada)^2\tP &= \lvb\Pa\rvb^4 + \lsb (2\lvb\Pa\rvb^2+1)\Pa^* a + \mathrm{h.c.}\rsb \\
			&+ \lvb\Pa\rvb^2(4\ada+1) + (aa\Pa^{*2}+\mathrm{h.c.}) \\
			&+ (2\ada a\Pa^* + \mathrm{h.c.}) + (\ada)^2
		\end{split}
	\end{align}
\end{subequations}
where we have defined the projection operator
\begin{equation}
	\Pa = \ag\Pg + \ae\Pe,
\end{equation}
with $\lvb\Pa\rvb^n = \lvb\ag\rvb^n\Pg+\lvb\ae\rvb^n\Pe$. Using these results, we apply the transformation $\tP$ to the Hamiltonian $H\trans{\atD} = H_s\trans{\atD}+H_d\trans{\atD}$ to obtain
\newcommand{\hc}{\mathrm{h.c.}}
\begin{equation}
	\begin{split}
		\raisetag{155pt}
		H\trans{\atD\atP} &= \hbar\Delta_{rm}' \lvb\Pa\rvb^2 + \hbar\zeta\lvb\Pa\rvb^4\sz \\
		&\quad + \hbar(\Pa^*\epm+\hc) \lp 1+\frac{\lambda^2\sz}{2}\rp \\
		&\quad + \hbar\lsb \wa+\chi + 2(\chi+\zeta)\lvb\Pa\rvb^2\rsb\frac{\sz}{2} \\
		&\quad + \hbar\lsb \Delta_{rm}'\Pa + (\chi+\zeta)\Pa\sz + 2\zeta\lvb\Pa\rvb^2\Pa\sz \vphantom{\frac{\lambda^2}{2}} \right. \\
		&\qquad\qquad  + \left. \epm\lp1+\frac{\lambda^2\sz}{2}\rp\rsb\ad + \hc \\
		&\quad + \hbar\lsb \Delta_{rm}' + \lp \chi+\zeta\ada+4\zeta\lvb\Pa\rvb^2\rp\sz\rsb\ada \\
		&\quad + \hbar\zeta \ad\ad\Pa^{2} + \hc + 2\hbar\zeta\ad\ada\Pa\sz+\hc
	\end{split}
\end{equation}
with $\Delta_{rm}' = \Delta_{rm} + \zeta$. Taking into account the time-dependence of $\tP$, the transformed Hamiltonian reads
\begin{equation}
	\label{eqn:H_DRP}
	H\trans{\atD\bar\atP} = H\trans{\atD\atP} - (i\hbar \dot \Pi_\alpha \ad + \mathrm{h.c.}) + \hbar\ImaginaryPart[\dot\Pi_\alpha\Pi_\alpha^*],
\end{equation}
the bar on the superscripts indicating that time dependence of the transformation is taken into account explicitly.

We also apply this transformation to the dissipative terms of the dispersive master equation \eqref{eqn:ExactMasterEquation_D}. For the first term of second line ($\kappa$ term), keeping up to order $\lambda^2$, we get
\begin{equation}
	\label{eqn:polaron_dissip_a}
	\begin{split}
		\raisetag{32pt}
		&\sD[a\trans{\atD\atP}]\rho \approx \sD\lsb a\trans{\atP}\lp 1+\frac{\lambda^2\sz}{2}\rp\rsb\rho =
		\sD\lsb a\lp 1+\frac{\lambda^2\sz}{2}\rp\rsb\rho \\
		& \qquad + \com{\sz}{\rho} \frac{\ad}{2} (\beta+\lambda\Pa) + \hc \\
		& \qquad + \frac{1}{4}[\lvb\beta\rvb^2 + \lambda^2 (\ne-\ng)] \sD[\sz]\rho \\
		& \qquad - i \frac{\ImaginaryPart[\ag\ae^*]}{2} \com{\sz}{\rho} \\
		& \qquad - i \frac{1}{2} \com{-i\Pa\lp1+\lambda^2\sz\rp \ad + \mathrm{h.c.}}{\rho} + \order{\lambda^4}.
	\end{split}
\end{equation}

We then get for the $\gamma$ and $\kappa_\gamma$ terms
\begin{subequations}
	\label{eqns:polaron_dissip_gamma}
	\begin{align}
		\label{eqn:polaron_dissip_gamma}
		\begin{split}
			\raisetag{12pt}
			& \sD\lsb\sm\trans{\atP}\lcb 1-\lambda^2\lp\ada+\frac12\rp\rcb\trans{\atP}\rsb\rho = \\
			& \quad \lsb 1-2\lambda^2\lp\ne+\frac12\rp\rsb\sD[\sm\trans{\atP}]\rho \\
			& \quad -\ae\lambda^2 D(\beta)\sm(\ad\rho+\rho\ad)\sp D^\dag(\beta) + \hc \\
			& \quad + 2\ae\lambda^2 \rho\ad\sp\sm + \hc \\
			& \quad -\lambda^2\lcb D(\beta)\sm\rho\sp\ada D^\dag(\beta) - \rho\ada\sp\sm \rcb +\hc \\
			& \quad -i \lambda^2\com{i\ae\Pe\ad + \hc}{\rho} + \order{\lambda^4},
		\end{split} \\
		\label{eqn:polaron_dissip_kappa_gamma}
		\begin{split}
			\raisetag{12pt}
			& \sD[a\trans{\atP}\sz]\rho = \sD[a\sz]\rho + \frac{\lvb\beta\rvb^2}{4}\sD[\sz]\rho - \frac{i\ImaginaryPart[\ag\ae^*]}{2}\com{\sz}{\rho} \\
			& \qquad\qquad\qquad + \lsb \sz\Pa\rho\ad\sz - \rho\ad\Pa \rsb + \hc \\
			& \qquad\qquad\qquad -i\frac12\com{-i\ad\Pa + \hc}{\rho},
		\end{split}
	\end{align}
\end{subequations}
and, for the $\gphi$ term,
\begin{equation}
	\label{eqn:polaron_dissip_gphi}
	\begin{split}
		\raisetag{12pt}
		&\sD\lsb\sz\lcb 1-2\lambda^2\lp\ada+\frac12\rp\rcb\trans{\atP}\rsb\rho = \\
		& \quad \lsb1-2\lambda^2\lp n_e+n_g+1\rp\rsb \sD[\sz]\rho \\
		& \quad - 2\lambda^2\lsb \ada \sD[\sz]\rho + \hc\rsb \\
		& \quad - 2\lambda^2\lsb (\sD[\sz]\rho) \ad \Pa + \hc \rsb \\
		& \quad - 2\lambda^2\lsb \ad\Pa \sD[\sz]\rho + \hc \rsb + \order{\lambda^4}.
	\end{split}
\end{equation}
Finally, for $\gamma_{\pm\Delta}$, we have
\begin{subequations}
	\label{eqns:polaron_dissip_gamma_pm_delta}
	\begin{align}
		\begin{split}
			\raisetag{30pt}
			\sD[a\trans{\atP}\sp\trans{\atP}]\rho &= \ng\sD[\sp\trans{\atP}]\rho + \sD[\sp\trans{\atP}a]\rho \\
			& \quad + \ag(D^\dag(\beta) \sp\rho\sm \ad D(\beta) - \rho\ad\Pg) + \hc \\
			& \quad -i \frac12 \com{-i\ag\ad\Pg + \hc}{\rho}
		\end{split} \\
		\begin{split}
			\raisetag{16pt}
			\sD[\ad\trans{\atP}\sm\trans{\atP}]\rho &= \ne\sD[\sm\trans{\atP}]\rho + \sD[\sm\trans{\atP}\ad]\rho \\
			&\quad + \ae\lsb D(\beta)\ad\sm\rho\sp D^\dag(\beta) - \rho\ad\Pe\rsb + \hc \\
			&\quad -i \frac12\com{-i\ae\ad\Pe+\hc}{\rho}.
		\end{split}
	\end{align}
\end{subequations}
The last line of Eqs~\eqref{eqn:polaron_dissip_a}, \eqref{eqns:polaron_dissip_gamma} and \eqref{eqns:polaron_dissip_gamma_pm_delta} act like a drive Hamiltonian. We will be able to cancel them with $\ae$ and $\ag$ given by Eq~\eqref{eqn:Condition_alphas}. In the above expressions, the quantities $n_{\g} = \lvb\alpha_{\g}\rvb^2$ and $n_{\e} = \lvb\alpha_{\e}\rvb^2$ are the number of photons when the qubit in the ground or excited state, and we have $\sD[\sm\trans{\atD}] = \sD[D(\beta)\sm]$ and $\sD[\sp\trans{\atD}] = \sD[D^\dag(\beta)\sp]$. 

If we put all the results of this section together, we can write the polaron-frame master equation, which is given by applying the polaron transform on Eq.\eqref{eqn:ExactMasterEquation_D}. The result is given by combining the results from Eqs~(\ref{eqn:H_DRP}-\ref{eqns:polaron_dissip_gamma_pm_delta})
\begin{equation}
	\label{eqn:EquationMaitresse_DRP}
	\begin{split}
		\dot\crho\trans{\atD\atP} &= -i \com{H\trans{\atD\bar\atP}}{\crho\trans{\atD\atP}} \\
		&\quad + \kappa\sD[a\trans{\atP}(1+\lambda^2\sz/2)]\crho\trans{\atD\atP} + \gamma_\kappa\sD[\sm\trans{\atP}]\crho\trans{\atD\atP} \\
		&\quad + \gamma\sD\lsb\sm\trans{\atP}\lcb 1-\lambda^2\lp\ada+\frac12\rp\rcb\trans{\atP}\rsb\crho\trans{\atD\atP} \\
		&\quad + \kappa_\gamma \sD[a\trans{\atP}\sz]\crho\trans{\atD\atP} \\
		&\quad + \frac{\gphi}{2} \sD\lsb \sz\lcb 1-2\lambda^2\lp\ada+\frac12\rp\rcb\trans{\atP}\rsb\crho\trans{\atD\atP} \\
		&\quad + \gamma_\Delta \sD[\ad\trans{\atP}\sm\trans{\atP}]\crho\trans{\atD\atP} + \gamma_{-\Delta}\sD[a\trans{\atP}\sp\trans{\atP}]\crho\trans{\atD\atP}
	\end{split}
\end{equation}

\subsection{Reduced master equation} % (fold)
\label{sub:a_reduced_master_equation}
In this section, we trace the transformed master equation Eq.~\eqref{eqn:EquationMaitresse_DRP}  over the resonator states to obtain an effective master equation for the qubit only.  This is done by first expressing the total density matrix in the polaron-transformed frame as
\begin{equation}
	\crho\trans{\atD\atP} = \sum_{n,m=0}^{\inf} \sum_{s,s'\in\{\e,\g\}} \crho\trans{\atD\atP}\melem{n,m,s,s'} \ket{n,s}\bra{m,s'}.
\end{equation}
Since our goal is to obtain the effective equation in the original non-polaron transformed frame, we write the reduced qubit density matrix in this frame as 
\begin{equation}
	\begin{split}
		\qrho\trans{\atD} &= \trace[r]{\tP\crho\trans{\atD\atP}\tP^\dag} = \sum_{s,s'\in\{\g,\e\}} \qrho\trans{\atD}\melem{s,s'} \ket{s}\bra{s'}
	\end{split}
\end{equation}
with
\begin{equation}
	\qrho\trans{\atD}\melem{s,s} \equiv \crho\trans{\atD\atP}\melem{0,0,s}, \qquad \qrho\trans{\atD}\melem{\e,\g} = \sum_{n,m=0}^\inf \lambda\trans{\atD\atP}\melem{n,m,m,n},
\end{equation}
where we have defined
\begin{equation}
	\crho\trans{\atD\atP}\melem{i,j,s} = \trace[r]{\ad^j a^i \crho\trans{\atD\atP}\melem{s,s}},
\end{equation}
with $\{s,s'\}\in\{\g,\e\}$, $\lambda\trans{\atD\atP}\melem{n,m,p,q} = \crho\trans{\atD\atP}\melem{n,m,\e,\g} d_{p,q}e^{-i\ImaginaryPart[\ag\ae^*]}$, and $d_{p,q} = \braketop{p}{D[\am]}{q}$ is the matrix element of the displacement operator in the number basis.

To obtain the master equation for $\qrho\trans{\atD}$, we simply find the equation of motion for the matrix elements of $\crho\trans{\atD\atP}$. More precisely, we will look at the equation of motion for $\crho\trans{\atD\atP}\melem{i,j,s}$. 

\begin{widetext}
\begin{equation}
	\label{eqn:matrix_element_nme}
	\begin{split}
		\dot \crho\trans{\atD\atP}\melem{n,m,\e} &= -i\lsb(\Delta_{rm}'+\chi+4\zeta\ne)(n-m) + \zeta(n^2-m^2)\rsb\crho\trans{\atD\atP}\melem{n,m,\e} - 2i\zeta(n-m)\crho\trans{\atD\atP}\melem{n+1,m+1,\e} \\
		&\quad -i\zeta\ae^2 \lsb2n\crho\trans{\atD\atP}\melem{n-1,m+1,\e} + n(n-1)\crho\trans{\atD\atP}\melem{n-2,m,\e}\rsb +i\zeta\ae^{*2} \lsb2m\crho\trans{\atD\atP}\melem{n+1,m-1,\e} + m(m-1)\crho\trans{\atD\atP}\melem{n,m-2,\e}\rsb \\
		&\quad -2i\zeta\ae \lsb (2n-m)\crho\trans{\atD\atP}\melem{n,m+1,\e} + n(n-1)\crho\trans{\atD\atP}\melem{n-1,m,\e} \rsb +2i\zeta\ae^* \lsb (2m-n)\crho\trans{\atD\atP}\melem{n+1,m,\e} + m(m-1)\crho\trans{\atD\atP}\melem{n,m-1,\e} \rsb \\
		&\quad -\lcb \lsb\kappa_\gamma+\kappa(1+\lambda^2) + \gamma_\Delta - 2\gamma\lambda^2\rsb \frac{n+m}{2} + \gamma\lsb 1-2\lambda^2\lp\ne+\frac12\rp\rsb + \gamma_\kappa + \gamma_\Delta\ne \rcb \crho\trans{\atD\atP}\melem{n,m,\e} \\
		&\quad -\lp\frac{\gamma_\Delta}{2}-2\gamma\lambda^2\rp (\ae\crho\trans{\atD\atP}\melem{n,m+1,\e}+\ae^*\crho\trans{\atD\atP}\melem{n+1,m,\e}) -(\gamma_\Delta-2\gamma\lambda^2) \crho\trans{\atD\atP}\melem{n+1,m+1,\e} \\
		&\quad + \gamma_{-\Delta} \trace[r]{D(\beta)\ad^m a^n D^\dag(\beta) (\ng \crho\trans{\atD\atP}\melem{\g,\g} + \ag  \crho\trans{\atD\atP}\melem{\g,\g}\ad + \ag^* a \crho\trans{\atD\atP}\melem{\g,\g} + a \crho\trans{\atD\atP}\melem{\g,\g}\ad)}
	\end{split}
\end{equation}
\end{widetext}
From this equation, we see that the only way the element $\crho\trans{\atD\atP}\melem{0,0,e}$ depends on the other elements is through the two last lines. Moreover, the only way the elements $n,m\neq0$ can be populated from an element $i<n,j<m$ is through the second, third and last lines. The rates at which these mechanisms act are of the order $\zeta\lvb\ae\rvb^2=\zeta\ne$ and $\gu \equiv \gamma_{-\Delta}\ng$. On the other side, these elements decay more quickly than the $0,0$ element because of the $\kappa$ term, which we assume is dominant compared to $\kappa_\gamma$, $\gamma_\Delta$ and $\gamma\lambda^2$. If the conditions $\ne \ll n_\kappa \equiv \kappa/\zeta$ and $\gu \ll \kappa$ are satisfied, we can assume there is no significant population of the $n,m\neq0$ matrix elements. We will have a similar equation for $\dot \crho\trans{\atD\atP}\melem{n,m,\g}$, with the conditions being $\ng \ll n_\kappa$ and $\gd \ll \kappa$, where $\gd$ is defined at Eq.~\eqref{eqn:Gamma_Down}. If these conditions are fulfilled, we can reduce the above equation and that for the $\g$ component to
\begin{subequations}
	\begin{align}
		\label{eqn:reduced_matrix_element_ee}
		\dot\qrho\trans{\atD}\melem{\e,\e} &= -\gd \qrho\trans{\atD}\melem{\e,\e} + \gu \qrho\trans{\atD}\melem{\g,\g} \\
		\label{eqn:reduced_matrix_element_gg}
		\dot\qrho\trans{\atD}\melem{\g,\g} &= -\gu \qrho\trans{\atD}\melem{\g,\g} + \gd \qrho\trans{\atD}\melem{\e,\e}.
	\end{align}
\end{subequations}

On the other hand, the off-diagonal elements of the reduced qubit density matrix involve off-diagonal elements of the resonator density matrix and  we must consider the equation of motion for all the terms $\lambda\trans{\atD\atP}\melem{n,m,p,q}$
\begin{equation}
	\label{eqn:Mouv_Lambda1}
	\begin{split}
		\raisetag{32pt}
		\dot\lambda\trans{\atD\atP}\melem{n,m,p,q} &= \dot\crho\trans{\atD\atP}\melem{n,m,\e,\g} d_{p,q} e^{-i\ImaginaryPart[\ag\ae^*]} - i\partial_t\ImaginaryPart[\ag\ae^*]\lambda\trans{\atD\atP}\melem{n,m,p,q} \\
		&\quad + \dot\am\sqrt{p}\lambda\trans{\atD\atP}\melem{n,m,p-1,q} - \dot\am^* \sqrt{q}\lambda\trans{\atD\atP}\melem{n,m,p,q-1} \\
		&\quad - \frac12 \partial_t(\am\am^*)\lambda\trans{\atD\atP}\melem{n,m,p,q}.
	\end{split}
\end{equation}
If we do this and compute $\dot\crho\trans{\atD\atP}\melem{n,m,\e,\g}$ according to \eqref{eqn:EquationMaitresse_DRP}, we get an equation that can be reduced only to the element $\crho\trans{\atD\atP}\melem{0,0,\e,\g}$ in the conditions stated above $(\ne,\ng\ll n_\kappa)$. Considering that only the $0,0$ element is ever populated significantly, the equation of motion is then 
\begin{equation}
	\label{eqn:reduced_matrix_element_eg}
	\begin{split}
		\dot\qrho\trans{\atD}\melem{\e,\g} &= \dot\lambda\trans{\atD\atP}\melem{0,0,0,0} = -i(\omega_a\trans{\atP}+\partial_t\ImaginaryPart[\ag\ae^*])\qrho\trans{\atD}\melem{\e,\g}\\
		&\quad  - \lsb \frac{\gu+\gd}{2} + \lp\gphi_{\rm eff}\trans{\atP} + \frac12\partial_t(\am\am^*)\rp\rsb\qrho\trans{\atD}\melem{\e,\g}, 
	\end{split}
\end{equation}
with 
\begin{subequations}
	\begin{align}
		\begin{split}
			\omega_a\trans{\atP} &= \wa + \chi + \RealPart\lsb \epm\lp\beta^*+\frac{\lambda^2\mu^*}{2}\rp\rsb \\
			&\quad - \zeta (\ne^2+\ng^2) + (\kappa+\kappa_\gamma)\ImaginaryPart[\ag\ae^*]
		\end{split} \\
		\begin{split}
			\gphi_{\rm eff.}\trans{\atP} &= \gphi[1-2\lambda^2(\ne+\ng+1)] \\
			&\quad + (\kappa+\kappa_\gamma)\frac{\lvb\beta\rvb^2}{2} + \frac{\kappa\lambda^2(\ne-\ng)}{2} + \frac{\gamma_\Delta}{2}.
		\end{split}
	\end{align}
\end{subequations}

Using the expression~\eqref{eqn:Condition_alphas} for $\alpha_{g(e)}$, we can combine the equations of motion for the reduced qubit density matrix $\qrho\trans{\atD}$ to find the reduced qubit master equation~\eqref{eqn:ReducedMasterEquation} with the frequency and rates given by Eqs.~\eqref{eqn:Delta_acDR}, \eqref{eqn:Gamma_Phieff}, \eqref{eqn:Gamma_Down} and \eqref{eqn:Gamma_Up}. 
% We emphasize that this master equation for the qubit only is in the non-displaced frame.

% subsection a_reduced_master_equation (end)
% section the_polaron_transformation (end)

\section{The effective qubit quantum trajectory equation} % (fold)
\label{sec:derivation_of_the_effective_qubit_quantum_trajectory}
The QTE is derived using linear quantum measurement theory \cite{gambetta:2005a,gambetta:2008a}.  The linear form of Eq.~\eqref{eqn:ConditionalMasterEquationFull} is
\begin{equation}
	\begin{split}
		\dot{\bar\crho}_J\trans{\atD} &= \sL\trans{\atD} \bar\crho_J\trans{\atD} + 2\sqrt{\kappa\eta} \bar\sM[I_\phi(1+\lambda^2\sz/2)]\bar\crho_J\trans{\atD} J \\
		& \quad + i\sqrt{\kappa\eta}\com{Q_\phi(1+\lambda^2\sz/2)}{\bar\crho_J\trans{\atD}} J,
	\end{split}
\end{equation}
where the bar means that the state is not normalized and the linear measurement superoperator is
\begin{equation}
	\bar\sM[c]\crho\trans{\atD} = (c\crho\trans{\atD} + \crho\trans{\atD} c)/2.
\end{equation}
Moving to the frame defined by Eq.~\eqref{eqn:PolaronTransformation} yields
\begin{equation}
	\label{eqn:conditional_reduced_1}
	\begin{split}
		\raisetag{12pt}
		\dot{\bar\crho}_J\trans{\atD\atP} &= \sL\trans{\atD\atP} \bar\crho_J\trans{\atD\atP} + \sqrt{\kappa\eta}\lsb a\lp 1+\frac{\lambda^2\sz}{2}\rp e^{-i\phi} \bar\crho_J\trans{\atD\atP} + \mathrm{h.c.} \rsb J\\
		&\quad + \sqrt{\kappa\eta} \lsb \RealPart[\tilde\beta_\phi] \bar\sM[\sz]\bar\crho_J\trans{\atD\atP} + \RealPart[\tilde\mu_\phi] \bar\crho_J\trans{\atD\atP} \rsb J \\
		&\quad + i\sqrt{\kappa\eta} \frac{\ImaginaryPart[\tilde\beta_\phi]}{2} \com{\sz}{\bar\crho_J\trans{\atD\atP}} J,
	\end{split}
\end{equation}
where we have defined
\begin{equation}
	\tilde\beta_\phi = \lp \beta+\frac{\lambda^2\mu}{2}\rp e^{-i\phi}, \qquad \tilde\mu_\phi = \lp \mu + \frac{\lambda^2\beta}{2}\rp e^{-i\phi}.
\end{equation}
As it should, for $\lambda^2=0$ the three equations above are of the same form as those obtained in Ref.~\cite{gambetta:2008a}.  

As before we now find the equations of motion for the coefficients $\bar\crho\trans{\atD\atP}\melem{n,m,\e}$, $\bar\crho\trans{\atD\atP}\melem{n,m,\g}$, and $\bar\lambda\trans{\atD\atP}\melem{n,m,p,q}$. For the $\bar\crho\trans{\atD\atP}\melem{n,m,\e}$ element, we find
\begin{equation}
	\begin{split}
		\dot{\bar\crho}\trans{\atD\atP}\melem{n,m,\e} &= \eqref{eqn:matrix_element_nme} + \sqrt{\kappa\eta}(\RealPart[\tilde\mu_\phi]+\RealPart[\tilde\beta_\phi]) \bar\crho\trans{\atD\atP}\melem{n,m,\e} J \\
		&\quad + \sqrt{\kappa\eta}\lp1+\frac{\lambda^2}{2}\rp \lsb e^{-i\phi} \bar\crho\trans{\atD\atP}\melem{n+1,m,\e} + e^{i\phi} \bar\crho\trans{\atD\atP}\melem{n,m+1,\e}\rsb J,
	\end{split}
\end{equation}
with a similar equation for the $\bar\crho\trans{\atD\atP}\melem{n,m,\g}$ component. For the $\bar\lambda\trans{\atD\atP}\melem{n,m,p,q}$ component, we find
\begin{equation}
	\begin{split}
		\raisetag{20pt}
		&\dot{\bar\lambda}\trans{\atD\atP}\melem{n,m,p,q} = \eqref{eqn:Mouv_Lambda1} + \sqrt{\kappa\eta}\lsb \vphantom{\frac{\lambda^2}{2}} (\RealPart[\tilde\mu_\phi] + i\ImaginaryPart[\tilde\beta_\phi])\bar\lambda\trans{\atD\atP}\melem{n,m,p,q} \right. \\
		& + \lp1+\frac{\lambda^2}{2}\rp \sqrt{n+1} e^{-i\phi}\bar\lambda\trans{\atD\atP}\melem{n+1,m,p,q} \\
		& \left. +\lp1-\frac{\lambda^2}{2}\rp \sqrt{m+1}e^{i\phi}\bar\lambda\trans{\atD\atP}\melem{n,m+1,p,q} \rsb J.
	\end{split}	
\end{equation}
In these expressions, which are the contribution of the Linblad term $\sL\trans{\atD\atP}\bar\crho_J\trans{\atD\atP}$ of Eq.~\eqref{eqn:conditional_reduced_1}, the equation numbers refer to the RHS of the corresponding expressions. 

The added measurement and back-action operators in the evolution equations does not change the approximation used in the previous section. Therefore, in the same limits, we can consider that the only relevant components are $\bar\crho\trans{\atD\atP}\melem{0,0,\e}$, $\bar\crho\trans{\atD\atP}\melem{0,0,\g}$ and $\bar\lambda\trans{\atD\atP}\melem{0,0,0,0}$. We can then write
\begin{subequations}
	\begin{align}
		\begin{split}
			\raisetag{60pt}
			\dot{\bar\qrho}\trans{\atD}\melem{\e,\e} &= \eqref{eqn:reduced_matrix_element_ee} + \sqrt{\kappa\eta}\lsb \RealPart[\tilde\mu_\phi] + \RealPart[\tilde\beta_\phi]\rsb \bar\qrho\trans{\atD}\melem{\e,\e} J \\
			&\quad + \sqrt{\kappa\eta}\lp 1+\frac{\lambda^2}{2}\rp\lsb e^{-i\phi} \bar\crho\trans{\atD\atP}\melem{1,0,\e} + e^{i\phi} \bar\crho\trans{\atD\atP}\melem{0,1,\e}\rsb J
		\end{split} \\
		\begin{split}
			\raisetag{44pt}
			\dot{\bar\qrho}\trans{\atD}\melem{\g,\g} &= \eqref{eqn:reduced_matrix_element_gg} + \sqrt{\kappa\eta}\lsb \RealPart[\tilde\mu_\phi] - \RealPart[\tilde\beta_\phi]\rsb \bar\qrho\trans{\atD}\melem{\g,\g} J \\
			&\quad + \sqrt{\kappa\eta}\lp 1-\frac{\lambda^2}{2}\rp\lsb e^{-i\phi} \bar\crho\trans{\atD\atP}\melem{1,0,\g} + e^{i\phi} \bar\crho\trans{\atD\atP}\melem{0,1,\g}\rsb J,
		\end{split} \\
		\begin{split}
			\dot{\bar\qrho}\melem{\e,\g}\trans{\atD} &= \eqref{eqn:reduced_matrix_element_eg} + \sqrt{\kappa\eta} \lsb \RealPart[\tilde\mu_\phi] + i \ImaginaryPart[\tilde\beta_\phi]\rsb \bar\qrho\trans{\atD}\melem{e,g} J,
		\end{split}
	\end{align}
\end{subequations}
and it is possible to construct a reduced linear QTE for the qubit in the dispersive frame
\begin{equation}
	\begin{split}
		\dot{\bar\qrho}\trans{\atD}_J &= \sL\trans{\atD}\bar\qrho\trans{\atD}_J + \sqrt{\kappa\eta} \RealPart[\tilde\beta_\phi] \bar\sM[\sz]\bar\qrho\trans{\atD}_J J \\
		&\quad + i\frac{\sqrt{\kappa\eta}\ImaginaryPart[\tilde\beta_\phi]}{2}\com{\sz}{\bar\qrho\trans{\atD}_J} J + \sqrt{\kappa\eta}\RealPart[\tilde\mu_\phi]\bar\qrho\trans{\atD}_J J.
	\end{split}
\end{equation}
Using Eq.~\eqref{eqn:homodyne_current} and normalizing, the above QTE gives Eq.~\eqref{eqn:conditional_qubit_master_equation} with measurement record given by Eq.~\eqref{eqn:measurement_record}. 

% section derivation_of_the_effective_qubit_quantum_trajectory (end)

\bibliographystyle{apsrev}

\end{document}